\RequirePackage[utf8]{inputenc}
\documentclass[jws,noconfig]{now-journal}
\usepackage{multirow}
\usepackage{dblfloatfix}
\usepackage{booktabs}
\usepackage{natbib}
\usepackage{todonotes}
\title{The Wisdom of the Few?\\``Supertaggers'' in Collaborative Tagging Systems}

\affil{Indiana Univeristy Cognitive Science Program}
\affil{Indiana University Department of Psychological \& Brain Sciences}
\affil{Indiana University School of Informatics and Computing}
\affil{Massachusetts General Hospital and Harvard Medical School:\\Division of Neurotherapeutics, Department of Psychiatry}
\author[1,2]{Lorince, Jared}
\author[4]{Zorowitz, Sam}
\author[1,3]{Murdock, Jaimie}
\author[1,2,3]{Todd, Peter M.}

\begin{document}
%

%
%
%
%
%

\maketitle

\keywords{Collaborative tagging, Folksonomy, Supertaggers}

\begin{abstract}
A folksonomy is ostensibly an information structure built up by the ``wisdom of the crowd'', but is the ``crowd'' really doing the work? Tagging is in fact a sharply skewed process in which a small minority of ``supertagger'' users generate an overwhelming majority of the annotations. Using data from three large-scale social tagging platforms, we explore (a) how to best quantify the imbalance in tagging behavior and formally define a supertagger, (b) how supertaggers differ from other users in their tagging patterns, and (c) if effects of motivation and expertise inform our understanding of what makes a supertagger. Our results indicate that such prolific users not only tag \emph{more} than their counterparts, but in quantifiably \emph{different} ways. Specifically, we find that supertaggers are more likely to label content in the long tail of less popular items, that they show differences in patterns of content tagged and terms utilized, and are measurably different with respect to tagging expertise and motivation. These findings suggest we should question the extent to which folksonomies achieve crowdsourced classification via the ``wisdom of the crowd'', especially for broad folksonomies like Last.fm as opposed to narrow folksonomies like Flickr.  
\end{abstract}

\section{Introduction}
\label{sec:intro}
In social tagging systems, users annotate content with freeform textual tags that can facilitate organization, sharing, and discovery of resources. Each instance of tagging is referred to as an annotation, and can be formally represented as a four element tuple (user-item-tag-time) indicating which user tagged which resource, the tag used, and the time of the annotation. Participation rates in these systems vary widely, from users who never tag to ``supertaggers'' who tag thousands of resources. This imbalance in contribution rates has important implications for how we interpret social tagging data, especially as most users are precisely that: users. They may use tags to search for or gain information about resources, but only some users actively contribute to the knowledge-generation process through tagging.

How effective or useful folksonomies are in general is not a topic we address here. Instead, our research questions the assumption that the ``crowd'' is at play in any meaningful way in collaborative tagging. Our results demonstrate that an overwhelming proportion of unique tagging is carried out by a minority of users, suggesting that the folksonomy does not necessarily represent the aggregated knowledge of its users, but is instead dominated by contributions from the few ``supertaggers'' among them.

Underlying this discrepancy is the fundamental issue of motivation --- why do users contribute to social tagging systems in the first place? A substantial literature has explored this topic in terms of why users tag in one manner rather than another \citep{Nov2010,Ames2007,Strohmaier2010}, but there is little work addressing what motivates some users to tag so much more than others. The differences we find between supertaggers and other users can be used to explore the motivational factors that may distinguish these two groups.

The relative importance that users place on tagging content versus other available activities is certainly a key factor here, and may explain variation in the relative contributions of supertaggers from one tagging system to another. On the social bookmarking site Delicious, for instance, tagging and organizing bookmarks is the principle use case for the service. However, on other services tagging is a secondary feature. Systems like Last.fm and Flickr incorporate tagging features, but their principal use cases (learning about and listening to music on Last.fm, photo sharing and discovery on Flickr) do not involve tagging. Many active users never make any substantive contribution to these systems' folksonomies. Such cases, in which tagging is a deliberate choice with costs of time and effort outside the primary use of a service, partly account for the lack of tagging participation we observe.  

In summary, the high-level question that interests us is this: \emph{How does the disproportionate contribution to the folksonomy by a small number of users change the interpretation of the presumed crowdsourced nature of tagging?} In other words, does the folksonomy truly represent the collective knowledge of its users? We approach this by exploring three research questions:
\begin{itemize}
	\itemsep0em
	\item RQ1: How do we most usefully quantify the imbalanced tagging contributions observed in collaborative tagging systems, and how do we formally define the term ``supertagger''?
	\item RQ2: How do observed patterns of tagging differ between supertaggers and other users? Do supertaggers simply tag \emph{more} or do they tag \emph{differently}?
	\item RQ3: What might be driving these differences? Can motivational or expertise effects, for instance, distinguish supertaggers from their counterparts, and how do such differences inform our interpretation of folksonomic data?
\end{itemize}

Expanding on an earlier version of this paper \citep{Lorince2014}\footnote{Aside from general refinements, such as improved visualizations of earlier analyses, the most notable additions to this paper compared to the earlier version are: the inclusion of additional datasets (the original paper only considered data from Last.fm); new analysis of consensus effects; and a much expanded section on tagging expertise, including two new expertise measures.} that analyzed tagging in Last.fm, here we address these questions across two additional large-scale tagging datasets from Delicious and Flickr. After presenting related work (Section~\ref{sec:related}) and an overview of the datasets (Section~\ref{sec:dataset}), we formalize our definition of supertaggers and illustrate their disproportionate tagging contribution in Section~\ref{sec:supertaggers} (RQ1). In Section~\ref{sec:analyses} we present our analyses of supertaggers' behaviors as compared to other users (RQ2). Next, in Section~\ref{sec:implications}, we explore RQ3 by examining if supertaggers differ from other users in terms of motivation and expertise. We conclude in Section~\ref{sec:conclusion} by synthesizing our results and discussing future avenues for work. 


Overall our findings demonstrate that a small proportion of users, the supertaggers, generate a disproportionate share of the tagging activity. This in and of itself may not be surprising, as long-tailed distributions in user activity (including, but not limited to tagging) on the web, are well-established. We also show, however, that their tagging patterns are quantifiably different than those of other users. This holds with respect to both the content they tag, most notably that supertaggers are more likely to label content in the long tail of less popular items, and the terms they use to tag it. Using established measures, as well as two novel methods, we also find that supertaggers show greater expertise and differing tagging motivations than other users. Precisely \emph{why} some users tag so much more than others may be partly accounted for by describer-like, as opposed to categorizer-like, tagging motivations (see Section~\ref{sec:motivation}), but remains a direction for future work discussed further in the conclusion.

\section{Related Work}
\label{sec:related}
\subsection{Folksonomies: From Individual Tagging Choices to Social Content Classification}
Thomas Vander Wal originally coined the term ``folksonomy'' in a 2004 listserv posting, describing it as a ``user-created bottom-up categorical structure'' that ``is the result of personal free tagging of information and objects (anything with a URL) for one's own retrieval.'' \citep{VanderWal2007}. Folksonomies are typically classified as broad, in which many users tag the same resources (e.g. music on Last.fm), or narrow, in which users typically tag their own content, such as a user's photos on Flickr \citep{VanderWal2005}.

Whereas many classification schemes are ``top-down'' taxonomies, a folksonomy is a ``bottom-up'' schema. In a taxonomy, a fixed set of pre-existing, often expert-generated, categories are used to classify resources. In a folksonomy, the vocabulary is unconstrained and comes from the users themselves, who may or may not be domain experts, bringing ``power to the people'' \citep{Quintarelli2005}. Many efforts have been made to infer taxonomies from folksonomies, synthesizing the advantages of controlled vocabulary and crowdsourced curation \citep{Kubek2010,Niepert2007}.

Due to their low economic costs of implementation and curation, folksonomies have been implemented in diverse domains, including Flickr (photos, \citealp{Nov2008,Nov2010}), Delicious (web bookmarks, \citealp{Golder2006}), Last.fm (music, \citealp{Lorince2013,Lorince2014}), and Bibsonomy (academic papers, \citealp{Hotho2006}). Reviews of many early social tagging systems can be found in \citet{Marlow2006} and \citet{Sen2006}. 

\subsection{Measuring Motivation in Tagging}
One factor that may differentiate supertaggers from other users, and may modulate levels of tagging in general, is tagging motivation: Different tagging goals may lead users to tag more or less. Though motivation in tagging behaviors has been operationalized in numerous ways, one prominent approach \citep{Korner2010,Korner2010a} characterizes users as either categorizers or describers. When tagging, categorizers use a limited vocabulary to construct a personal taxonomy conducive to later browsing of tagged content. In contrast, describers do not constrain their vocabulary; instead, they freely use a variety of informative keywords to describe items, facilitating later keyword-based search. \citet{Strohmaier2010} and \citet{Korner2010} present several metrics with which to classify users according to this dichotomy, discussed in Section~\ref{sec:implications}. Other researchers have developed taxonomies of tagging motivation that can be broadly mapped onto dimensions of sociality (are tags self- or socially-directed?) and function (are tags used for organization or communication?) \citep{Ames2007,Heckner2009}. Methods for identifying these motivations programmatically in large-scale datasets have yet to be developed, however.

\subsection{Measuring Expertise in Tagging}
Another important consideration for studying user contributions to a folksonomy is expertise. Inevitably, some annotations will provide more useful information about an item than others. Expert users presumably generate higher quality annotations on average. 

Though expertise has no single agreed-upon definition with respect to tagging, one noteworthy approach to expert detection is Spamming-Resistant Expertise Analysis and Ranking (SPEAR, \citealp{Yeung2009,Yeung2011}). SPEAR assigns an expertise score to users for each unique tag they use based on two principles. First, under a mutual reinforcement model, user expertise in a topic (as defined by a specific tag) is determined by the quality of items the user tags with that term, and item quality is in turn determined by the expertise of users who have tagged it. Second, users who tend to tag items earlier receive higher expertise scores, as they identify new, high quality resources sooner than others. In this way, SPEAR is adept at weeding out spammers, who tend to indiscriminately annotate items with tags. The use of a spam-robust expertise measure is important, as \citet{Wetzker2008} found an overwhelming majority of the most prolific taggers in a large folksonomy were spammers. 

We also consider a new expertise measure to supplement SPEAR. Because it evaluates users with respect to a given tag, SPEAR provides a useful measure of domain expertise, or knowledge of a particular topic. Our measure, on the other hand, is designed to evaluate general user expertise (i.e. across all annotations). In contrast to SPEAR, our approach evaluates users on an item-by-item basis, and assigns higher scores to users annotating an item in alignment with the consensus of annotations for that item. The details of this measure are discussed in Section~\ref{sec:expertise}, but it essentially asks if supertaggers are more likely to assign ``better'' tags to an item, where the quality of a tag is defined in terms of how much agreement there is across multiple users that it should be assigned to an item.

A third approach to expertise is inspired by classic research on the structure of mental categories \citep{Rogers2007,Rosch1976}, which suggests that linguistic consensus emerges around labels/words indexing categories of an intermediate level of abstraction. For example, people prefer basic-level terms (e.g. ``dog'') over super- and sub-ordinate terms (e.g., ``mammal'' and ``terrier'', respectively) to refer to an object. In contrast to the consensus, experts in a given domain tend to deviate from this verbal behavior reliably \citep{Tanaka1991}  by applying more specific (sub-ordinate) labels. \citet{Kubek2010} present a method based on conditional probabilities to automatically extract semantic taxonomies from folksonomic data, which allows us to to calculate a depth score for each tag in the resulting taxonomy. We can then determine if supertaggers tend to use more sub-ordinate terms, thereby showing evidence of greater expertise. 

A similar approach was taken by \citet{Fu2010}, who applied a Latent Dirichlet Allocation (LDA) model to a subset of items from Bibsonomy and found that resources tagged by experts, as determined by SPEAR, contained tags more predictive of topics as compared to those by non-experts. Such an approach is not applicable here, however, due to the fact that items in our datasets are either non-linguistic in nature (photos on Flickr, music on Last.fm) or not directly available (we have only arbitrary IDs for the URLs tagged in the Delicious dataset).
\section{Datasets}
\label{sec:dataset}
We performed our analyses on datasets from three different collaborative tagging systems, the social music site Last.fm, the photo-sharing site Flickr, and the social bookmarking tool Delicious. The targets of tagging are, respectively, music (users can tag artists, albums, or songs), photos (users tag the images they upload to the site), and web bookmarks (users save and tag links to webpages). 

The Flickr and Delicious datasets were collected by G\"{o}rlitz, Sizov, and Staab in 2006 and 2007\nocite{Gorlitz2008}, and are publicly available.\footnote{\texttt{http://www.uni-koblenz-landau.de/campus-koblenz/fb4/west/\\Research/DataSets/PINTSExperimentsDataSets}} These datasets consist exclusively of annotation data (i.e. tuples in the form user-resource-tag-date). 

The Last.fm dataset, on the other hand, was collected by the current authors. It is an expanded version of that first presented in \citet{Lorince2013} and later analyzed in \citet{Lorince2014}. 

\subsection{Crawling Methodology}
\label{sec:crawling}
We crawled the Last.fm data in 2013 with a combination of API queries and HTML scraping of users' publicly available profile pages. We did so on a user-by-user basis, such that we have the complete tagging history for every user in our data, but not necessarily the complete tagging history for any particular item (artist, album, or song). All temporal annotation data is at a monthly granularity, as users' profiles only list the month and year in which an item was tagged. Users were crawled by traversing the site's social network using a snowball sampling method (beginning with seed users having at least one friend). As such, we necessarily include only users with at least one friendship on the site, but we do not believe this is problematic for our analyses. This also means our sample includes users who have never tagged (unlike the Flickr and Delicious data). See \citet{Lorince2013} for further discussion of our crawling methods and its limitations. 

The Delicious dataset consists of an effectively random sample of users for whom complete tag histories were collected, much like our Last.fm data. For this data, however, annotations are recorded at an increased temporal resolution (seconds). The Flickr dataset was crawled at the tag level (i.e. the complete sets of annotations associated with an effectively random set of tags were collected). While this means that we cannot guarantee that any particular user's tag history is complete, we can assume that the number of tags ``missed'' is approximately equal across all users, such that the relative annotation counts over users accurately represent the true distribution. We base this assumption on (a) the sheer size of the sample (113 million annotations across 1.6 million tags), and (b) the fact that the distribution of annotations per user is generally consistent with the other two datasets. Further details of how the Flickr and Delicious datasets were collected can be found in \citet{Gorlitz2008}.

Overall the distributions of tagging activity are consistent both across datasets (as is particularly evident in Figure~\ref{fig:annoDists} of Section~\ref{sec:supertaggers}) and with previous work examining similar datasets (e.g. \citealt{Figueiredo2013}).

\begin{table}[h]
\begin{center}
\begin{tabular}{c|cccc}
\toprule
Dataset & Taggers & Tags & Resources & Annotations \\
\midrule
Last.fm & 521,780 & 1,029,091 & 4,477,593 & 50,372,895 \\
Flickr & 319,686 & 1,607,879 & 28,153,045 & 112,900,000 \\
Delicious & 532,924 & 2,481,108 & 17,262,475 & 140,126,555 \\
\bottomrule
\end{tabular}
\end{center}
\caption{Global tagging data summary.}
\label{tab:dataset}
\end{table}

\begin{table}[h]
\begin{center}
\begin{tabular}{c|cccc}
\toprule
Dataset & $A_{u}$ & $A_{t}$ & $A_{i}$  \\
\midrule
Last.fm & 7 (2-29) & 1 (1-4) & 2 (1-6) \\
Flickr & 41 (12-175) & 2 (1-10) & 3 (2-5) \\
Delicious & 41 (9-188) & 1 (1-4) & 2 (1-5)  \\
\bottomrule
\end{tabular}
\end{center}
\caption{Median number of annotations per user ($A_{u}$), tag ($A_{t}$), and item ($A_{i}$) across datasets. Interquartile range (25\textsuperscript{th} percentile - 75\textsuperscript{th} percentile) in parentheses.}
\label{tab:byUser}
\end{table}

\subsection{Data Summary}
\label{sec:datasummary}
Table~\ref{tab:dataset} gives an overview of the tagging data from all three tagging systems. Note that the ``taggers'' column reflects the total number of users with $\geq$1 annotation. While the crawling methods used in \citet{Gorlitz2008} are such that only users who have tagged are included, our Last.fm data also includes users who have never tagged. Unless otherwise noted, however, all analyses of Last.fm presented here are limited to the subset of users who have tagged at least once. Across all systems, an ``annotation'' refers to a given instance of a user assigning a particular tag to a particular item at a particular time. 

Even at this high level of description, substantial differences between these systems are apparent. Users clearly tag more overall on Flickr and Delicious (both with medians of 41 annotations per user) than on Last.fm (median of 7 annotations per user). Though the median numbers of annotations per item are similar across datasets, Last.fm has the greatest ratio of annotations to items tagged (11, versus 4 and 8 for Flickr and Delicious), suggesting a stronger trend towards popular, heavily tagged items. See Table~\ref{tab:byUser} for a summary of per-user, per-item, and per-tag median numbers of annotations (given the scale-free distributions of these measures, the mean does not accurately capture the central tendency of the data). 

These observations are consistent with the design of these systems. Following Vander Wal's \citeyearpar{VanderWal2005} terminology, Last.fm is a broad folksonomy in which many users tag the same, publicly available resources (i.e. multiple individuals tagging the same artists, albums, and songs), while Flickr is a narrow folksonomy, in which users predominantly tag their own photos.\footnote{In the vast majority of instances, photos can \emph{only} be tagged by the users who upload them, and in our data no single photo has been tagged by multiple users.} Delicious exists somewhere between these two extremes: On the one hand, users use the service to manage their own resources (in this case, Web bookmarks), much like Flickr. But on the other, multiple users can save and tag the same URL (either independently, or by exploring the bookmarks saved by other users on the site). Note that on Delicious and Last.fm, users receive the top five most popular tags for an item as recommendations when tagging it. On Last.fm (but not Delicious) users can browse the full tag distribution for an item, as well. On Flickr, where users tag only their own photos, such popularity-based recommendations are of course not possible.

The difference in sheer volume of tagging between the systems is also of note, with Last.fm having well less than half the total number of annotations of the other systems (despite having a comparable number of users). This is again consistent with how the systems are used. Tagging is a more central activity on both Flickr and Delicious, as users actively contribute and organize resources. Last.fm, on the other hand, is primarily used for music consumption, and tagging is generally speaking a non-primary activity for users.

\subsection{Supplemental Last.fm data}
\label{sec:supplemental}
For Last.fm, we crawled a total of nearly 1.9 million users, of whom about 28\% had tagged at least once. In addition to the tagging data, we recorded friendship relations, group memberships, loved/banned songs,\footnote{``Loving'' a track is roughly equivalent to favoriting a tweet, or other similarly-defined activities, while ``banning'' allows a user to indicate disliked items and exclude them from any recommendations by Last.fm.} and self-reported demographic data.  For a subset of our users, we also have collected full song listening (scrobble\footnote{Last.fm tracks users' listening for music recommendation purposes, and ``scrobble'' is the term for an instance of a user listening to a particular song at a particular time. }) histories. Table~\ref{tab:fmdata} summarizes the supplemental data collected.

\begin{table}[bh]
\begin{center}
\begin{tabular}{l|r}
\toprule
Total users & 1,884,597 \\
Friendship relations  & 24,320,919 \\ 
Total loved tracks & 162,788,213 \\
Total banned tracks & 23,321,347 \\
Unique groups & 117,663 \\
\midrule
Users with scrobbles recorded & 73,251 \\
Total scrobbles & 1,181,674,857 \\
Unique items scrobbled & 32,864,795 \\ 
\bottomrule
\end{tabular}
\end{center}
\caption{Supplemental data summary for Last.fm}
\label{tab:fmdata}
\end{table}

\section{Identifying ``Supertaggers'' and Measuring their Influence}
\label{sec:supertaggers}

\begin{figure}[t]
	\centering
	\includegraphics[width=8cm]{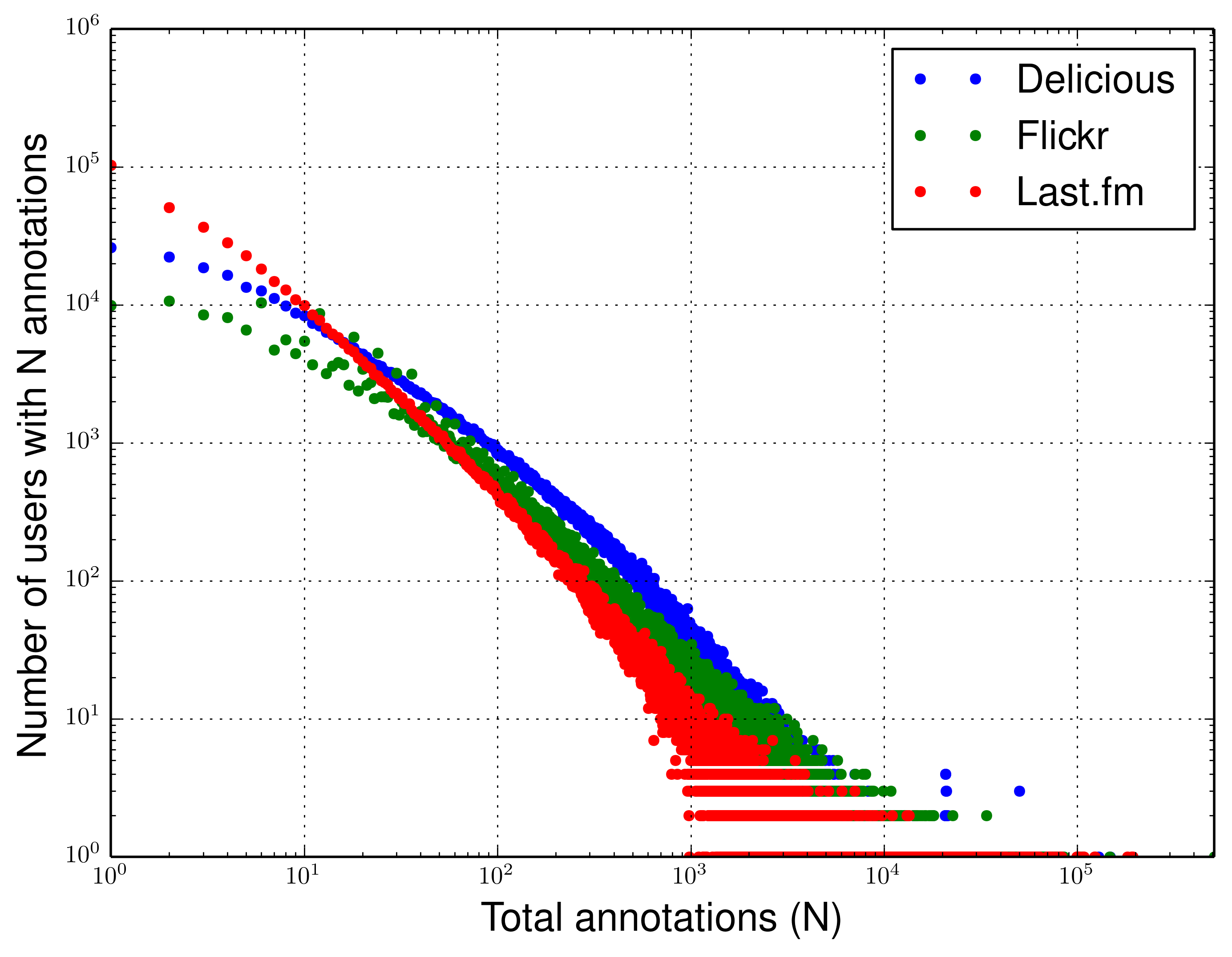}
	\caption{Frequency distributions of per-user annotation counts.}
	\label{fig:annoDists}	
\end{figure}

Figure~\ref{fig:annoDists} presents the distribution of per-user annotation counts for our three datasets in a traditional manner. For a given number of annotations on the x-axis, the corresponding y-axis value indicates how many users have generated that many total annotations. Plotted on a log-log scale, the distributions take roughly linear forms consistent with long-tailed, power-law-like distributions.\footnote{We do not examine the precise mathematical form of the distributions, as it is not relevant to our analyses.} Though Flickr is more variable for lower annotation counts than Last.fm and Delicious, the distributions generally decrease monotonically with increasing annotation counts, indicating that users with relatively small numbers of annotations are much more common in all three services.

\begin{figure}[t]
	\centering
	\includegraphics[width=8cm]{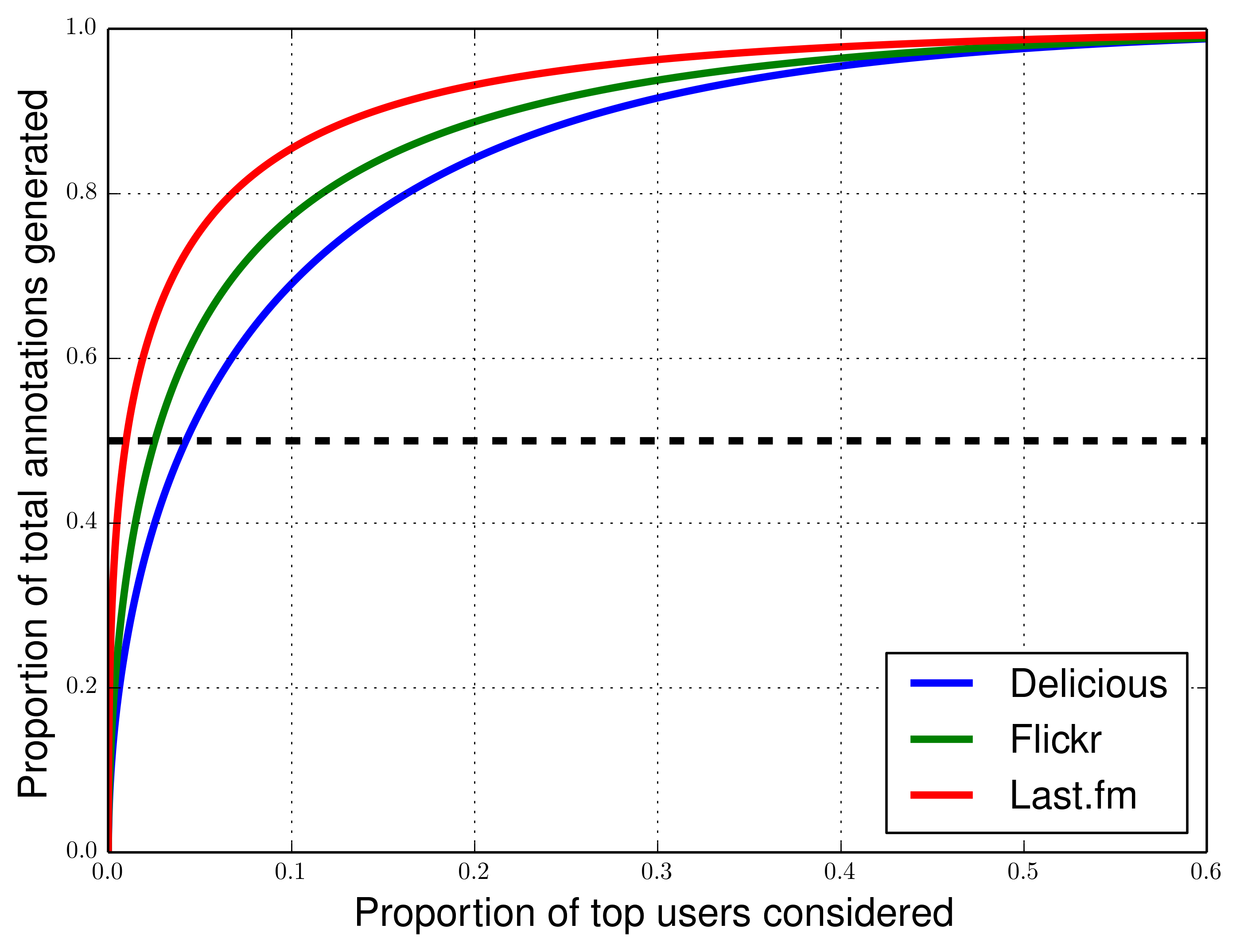}
	\caption{Proportions of total annotations generated by the most prolific taggers as a function of the proportion of top users considered. The dashed line shows the threshold used to identify supertaggers.}
	\label{fig:pareto}	
\end{figure}

These long-tailed distributions make it clear that there exist a relatively small number of prolific users generating many annotations and a large number of users generating only a few annotations. But to show exactly how pronounced this pattern is, we plot in Figure~\ref{fig:pareto} the proportion of total annotations generated by the most prolific taggers against the proportion of top taggers considered (i.e. the proportion $Y$ of annotations generated by the proportion $X$ of top taggers, ranked by total number of annotations). 

This representation highlights just how skewed these distributions are, even more so than predicted by the Pareto Principle \citep{Newman2005}, or 80-20 rule, under which we would expect 80\% of annotations to come from the top 20\% of users. Last.fm is the most extreme case, with 80\% of all annotations generated by less than 7\% of users, but both Flickr and Delicious show similarly skewed patterns, with approximately 12\% and 16\% of users, respectively, responsible for 80\% of all annotations. These findings are corroborated by calculating the Gini coefficient, a measure of income inequality:

\begin{equation*}
G = \frac{2\sum_{i=1}^{n} iy_{i}}{n\sum_{i=1}^{n} y_{i}} - \frac{n+1}{n}
\end{equation*}

\noindent where values of $y$ are individuals' ``wealth'' (here, their total numbers of annotations), indexed by $i$ in non-decreasing order ($y_{i}\leq y_{i+1}$), and $n$ is the total number of individuals. Values range from 0, indicating total equality (all individuals have equal wealth) to 1, total inequality (one individual has all the wealth). For all three datasets, we find high values of the Gini coefficient (0.806 for Delicious, 0.847 for Flickr, and 0.898 for Last.fm), indicating a small number of users performing most of the tagging. As expected, the effect is most pronounced on Last.fm. 

Our analyses do not reveal the existence of a clear split that naturally divides users into supertaggers and non-supertaggers, so an a priori definition of supertaggers based on annotation counts is necessarily arbitrary. One option would be to echo the Pareto principle, considering supertaggers to be the top 20\% of users, but under this definition over 90\% of tagging activity would be from supertaggers, making it difficult to compare the aggregated activity of supertaggers versus non-supertaggers. Instead, we elected to split the data in half with respect to annotations, allowing us to compare equally-sized sub-folksonomies (in terms of total annotations) from supertaggers and non-supertaggers (although it does mean the number of users we classify as supertaggers is much smaller). We hope these analyses will show the value of considering more prolific taggers separately, and can lead to more precise methods for identifying supertaggers in future work. 

Thus, ~\emph{we formally define ``supertaggers'' as the topmost prolific taggers accounting for half of all annotations in each dataset.} This split is marked by the horizontal dashed line in Figure~\ref{fig:pareto}. Under this definition, 5,086 users (0.97\%) from Last.fm, 8,142 users (2.55\%) from Flickr, and 22,630 users (4.25\%) from Delicious are classified as supertaggers. These correspond to annotation thresholds (i.e. the number of annotations required to be a supertagger) of 1,457, 2,701, and 1,285, respectively.

We reiterate that the particular threshold used here is arbitrary, in that there is no special behavioral shift that occurs at this point. In fact, various behavioral measures show relatively smooth changes as we consider users with progressively more annotations, and as such, we present measures as a function of users' total annotation counts (rather than simply comparing averages for supertaggers and non-supertaggers) wherever possible. Nonetheless, in those analyses that directly compare the annotations of the two groups we have defined, the 50 percent split is both clearly interpretable (in that it compares ``normal'' users to the most prolific ones), and analytically convenient, as it normalizes all analyses of the sub-folksonomies such that the total number of annotations in each is constant.

	\begin{table*}[b!]
	\begin{center}
	\begin{tabular}{p{2cm}cccccc}
	\cmidrule(r){2-7}
	\multicolumn{1}{c}{} & \multicolumn{2}{c}{Delicious} & \multicolumn{2}{c}{Flickr} & \multicolumn{2}{c}{Last.fm} \\
	\multicolumn{1}{c}{} & $S$ & $\lnot S$ & $S$ & $\lnot S$ & $S$ & $\lnot S$\\
	\midrule
	Users & 22,630 & 510,294 & 8,142 & 311,544 & 5,086 & 516,694 \\
	Annotations & 70,062,323 & 70,064,232 & 56,449,589 & 56,450,411 & 25,185,082 & 25,187,811 \\
	Total Tags & 1,210,748 & 1,698,863 & 803,772  & 1,094,358 & 399,552 & 797,784 \\
	Unique Tags & 782,245 & 1,270,360 & 513,521 & 804,107 & 231,307 & 629,539\\
	Shared Tags & \multicolumn{2}{c}{428,503} & \multicolumn{2}{c}{290,251} & \multicolumn{2}{c}{168,245} \\
	Total Items & 8,039,337 & 11,516,472 & 10,339,003 & 17,814,042 & 2,992,045 & 2,515,069 \\
	Unique Items & 5,746,003 & 9,223,138 & 10,339,003 & 17,814,042 & 1,962,522 & 1,485,546 \\
	Shared Items & \multicolumn{2}{c}{2,293,334} & \multicolumn{2}{c}{0} & \multicolumn{2}{c}{1,029,523} \\
	\bottomrule
	\end{tabular}
	\end{center}
	\caption{Summary statistics. ``Total Tags'' represent all distinct tags used by each group, while ``Unique Tags'' are those tags appearing in only the supertagger folksonomy ($S$) or the non-supertagger folksonomy ($\lnot S$). ``Shared Tags'' are those tags used at least once in both $S$ and $\lnot S$. Corresponding counts are shown for items.}
	\label{tab:folkstats}
	\end{table*}

	\begin{table*}[b!]
	\begin{center}
	\begin{tabular}{p{2cm}cccccc}
	\cmidrule(r){2-7}
	\multicolumn{1}{c}{} & \multicolumn{2}{c}{Delicious} & \multicolumn{2}{c}{Flickr} & \multicolumn{2}{c}{Last.fm} \\
	\multicolumn{1}{c}{} & $S$ & $\lnot S$ & $S$ & $\lnot S$ & $S$ & $\lnot S$\\
	\midrule
	Annotations & 2,115 (1,597-3,227) & 36 (8-153) & 4,615 (3,417-7,315) & 38 (11-153) & 2,586 (1,860-4,457) & 6 (2-27) \\
	Tags & 427 (265-644) & 19 (6-60) & 326 (129-670)  & 8 (3-24) & 209 (97-405) & 4 (2-12) \\
	Items & 641 (428-981) & 14 (3-62) & 952 (629-1,499) & 16 (6-56) & 882 (483-1,652) & 3 (1-15)\\
	\bottomrule
	\end{tabular}
	\end{center}
	\caption{Per-user summary statistics. Shown are the median number of annotations, unique tags, and items tagged \emph{per user} in $S$ and $\lnot S$, across datasets. Interquartile range (25\textsuperscript{th} percentile - 75\textsuperscript{th} percentile) in parentheses.}
	\label{tab:userStats}
	\end{table*}

\section{Differences in Tagging Patterns}
\label{sec:analyses}
This section presents analyses comparing the tagging patterns of supertaggers to those of their non-supertagger counterparts. Except where otherwise noted, analyses for each dataset were performed on two ``sub-folksonomies'', one containing all annotations by the supertaggers, designated $S$, and the other containing all annotations for non-supertaggers, designated $\lnot S$. Summary measures for these groups across all datasets appear in Table~\ref{tab:folkstats}, and Table~\ref{tab:userStats} shows the relevant per-user medians of these values. We analyze differences in tagging behavior from three perspectives: 
\begin{itemize}
\itemsep0em
\item How similar is the vocabulary of supertaggers and non-supertaggers? (Section~\ref{sec:vocab})
\item How much does the content tagged by supertaggers and non-supertaggers overlap? (Section~\ref{sec:content})
\item How similarly do supertaggers and non-supertaggers tag particular content? (Section~\ref{sec:consensus})
\end{itemize}



\subsection{Variation in Tagging Vocabulary}
\label{sec:vocab}

\begin{figure*}[t!]
	\centering
	\includegraphics[width=18cm]{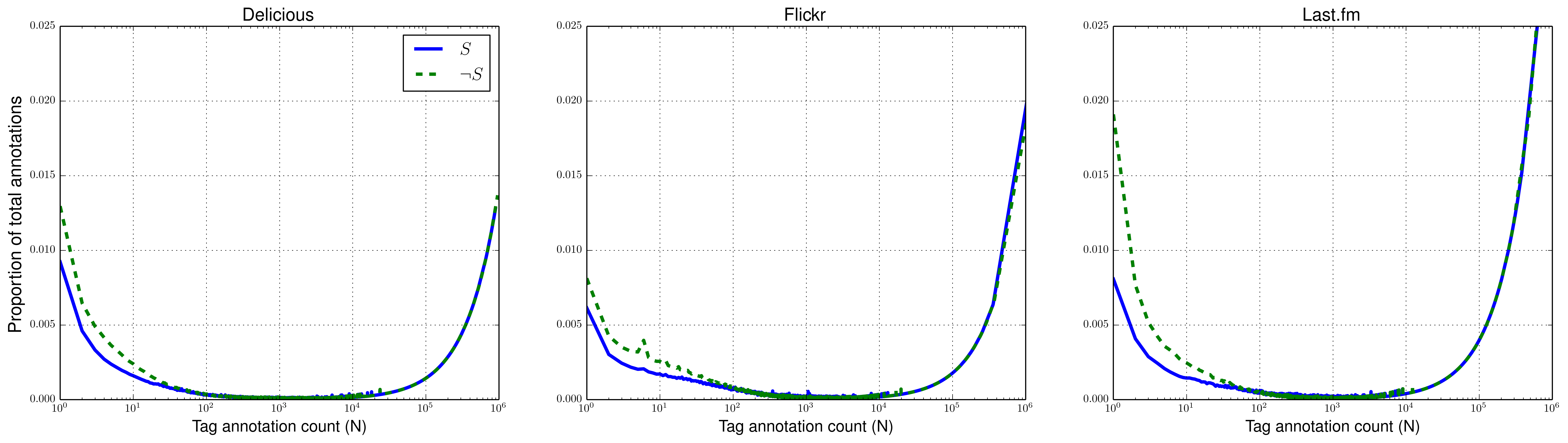}
	\caption{Distributions of tag usage for $S$ (blue) and $\lnot S$ (green) for all datasets. Each point indicates the proportion of total annotations within a sub-folksonomy allocated to tags that have been used $N$ total times.}
	\label{fig:tagFreqs}	
	\end{figure*}

We first ask how similar the tag vocabularies are between supertaggers and non-supertaggers, independent of the content they are annotating. While the set of non-supertaggers clearly employ a larger aggregate vocabulary (see Table~\ref{tab:folkstats}), the median number of unique tags per user is much greater for the supertaggers (see Table~\ref{tab:userStats}). However, both groups' vocabularies are largely shared, with most annotations coming from the set of shared tags occurring at least once in both $S$ and $\lnot S$ in each dataset (95\%, 88\%, and 93\%, respectively, for Delicious, Flickr, and Last.fm). This suggests the existence of many ``singletons'' -- tags used only once -- and other tags used only a small number of times. This is verified in Figure~\ref{fig:tagFreqs}, which compares the distributions tagging activity over tag popularity for $S$ and $\lnot S$. The plot shows the proportion of total annotations within each sub-folksonomy allocated to tags with a given total annotation count. Singletons and other low-frequency tags are clearly very common, and the u-shaped distributions show that, across both $S$ and $\lnot S$, and also across datasets, tagging is concentrated on a few popular tags and the many singleton tags (with proportionally little use of moderate-popularity tags). Note that on Delicious very popular and very rare tags show similar overall proportions of use indicating a more varied vocabulary, while on Flickr and Last.fm popular tags are proportionally more common.

\begin{figure*}[t!]
	\centering
	\includegraphics[width=18cm]{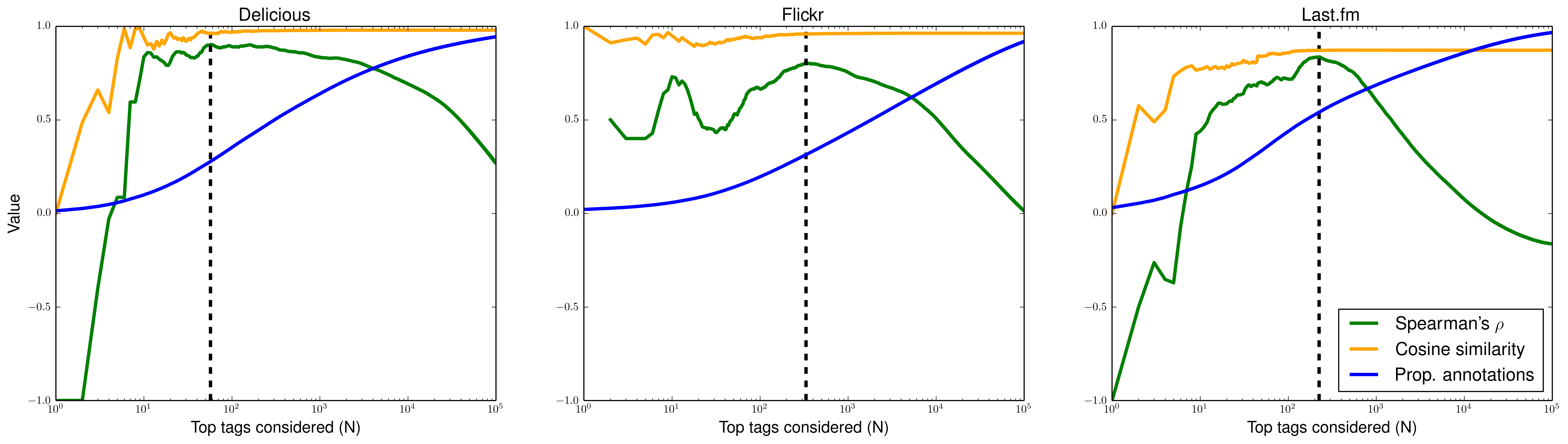}
	\caption{Spearman's $\rho$ and cosine similarity between $S$ and $\lnot S$ as a function of $N$, considering only the top $N$ most popular tags overall from each sub-folksonomy. Also plotted is the proportion of all annotations across the full folksonomy using the combination of the top $N$ tags from each group. The vertical dashed lines indicate the maximum values of $\rho$ and occur at $N=57$ (Delicious), $N=333$ (Flickr), and $N=224$ (Last.fm).}
	\label{fig:corrsTags}	
	\end{figure*}

To directly measure the similarity between the vocabularies of $S$ and $\lnot S$, we use two simple summary measures: the rank correlation, Spearman's $\rho$, of tags for each folksonomy (measuring how similar the rank order popularity is between the two vocabularies) and the cosine similarity between the two global tag vocabularies (i.e. calculated across vectors of the frequency of each tag in each of the two folksonomies). Considering all tags, we find low rank correlations of $\rho=-0.402$ for Delicious, $\rho=-0.256$ for Flickr, and $\rho=-0.219$ for Last.fm. In contrast, cosine similarity between $S$ and $\lnot S$ is high, with values of 0.979 for Delicious, 0.963 for Flickr, and 0.872 for Last.fm. These give rather opposing impressions of the distribution similarities, so it is informative to consider these measures for smaller subsets of the data. 

We calculated both measures for the top $N$ tags in both sub-folksonomies as a function of increasing $N$. For example, if $N=100$, we consider tag-frequency vectors for the top 100 most frequent tags in each sub-folksonomy (considered independently) and then calculate the rank correlation and cosine similarity of these vectors between $S$ and $\lnot S$. Tags that appear in $S$ but not $\lnot S$ (and vice versa) are assumed to have rank $N+1$ for the purposes of calculating the rank correlation, and frequency of zero for the cosine similarity calculation. This was repeated for $N$ from 1 to 100,000.

\begin{figure*}[b!]
	\centering
	\includegraphics[width=18cm]{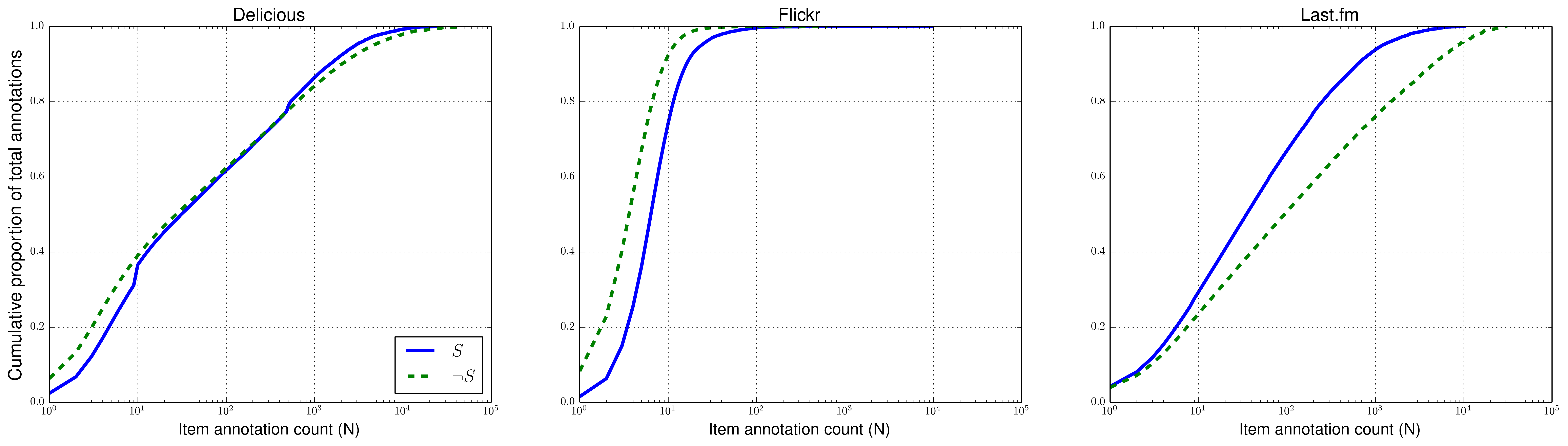}
	\caption{Distributions of item tagging for $S$ (blue) and $\lnot S$ (green) for all datasets. Each point indicates the \emph{cumulative} proportion of total annotations within a sub-folksonomy allocated to items that have been tagged at least $N$ total times.}
	\label{fig:itemFreqs}	
	\end{figure*}

Figure~\ref{fig:corrsTags} shows the results, additionally plotting the proportion of \emph{total} annotations (i.e. across the full folksonomy) generated by the combination of the top $N$ tags from $S$ and $\lnot S$ (e.g. for $N=100$, what is the sum proportion of total annotations from the top 100 tags from $S$ and the top 100 from $\lnot S$?).\footnote{Clearly, the top 100,000 tags  make up an overwhelming majority of total annotations. But note that the combination of the top $N$ tags from $S$ and $\lnot S$ contains more than $N$ unique tags overall. The proportion shown for$N=100,000$ tags, for example, corresponds to a total of 144,622, 155,702, and 160,470 unique tags, respectively, from the global folksonomies of Delicious, Flickr, and Last.fm.} Unsurprisingly, the rank correlation is noisy for small $N$, but across all datasets has a distinct peak (0.902 for Delicious, 0.804 for Flickr, and 0.836 for Last.fm) after which it decreases monotonically. Cosine similarity is also noisy for small $N$, but clearly stabilizes near the peak in $\rho$ (at approximately 0.96 for Delicious, 0.96 for Flickr, and 0.87 for Last.fm). 

These observations suggest a ``core'' vocabulary of common tags that $S$ and $\lnot S$ more or less agree on, the size of which is estimated by the index of the maximum value of $\rho$. The fact that cosine similarity stabilizes at approximately the same $N$ for which $\rho$ begins to decrease suggests that the cosine similarity between $S$ and $\lnot S$ is driven by the most popular tags (i.e. the top $N$ occurring before the peak in $\rho$). Thus we find core vocabulary sizes of the datasets are 57, 333, and 224, respectively for Delicious, Flickr, and Last.fm (indicated by the dashed vertical lines in Figure~\ref{fig:corrsTags}). These core vocabularies in turn account for 28\%, 31\%, and 54\% of all annotations (i.e. across $S$ and $\lnot S$) from the datasets. A possible explanation of the higher percentage observed for Last.fm is the existence of a relatively well-defined, constrained set of canonical music genres (``rock'', ``jazz'', ``classical'', and so on) that are common in music tagging. There is not an obvious analog to these popular categories with respect to photos (Flickr) or web bookmarks (Delicious). Running contrary to this reasoning, is the fact that on on Flickr there is strong agreement as to the most popular tags (as evidenced by relatively high cosine similarity and $\rho$ for low $N$). Despite this agreement, a relatively lower proportion of annotations come from these agreed-upon popular tags, suggesting that the ``vocabulary'' on Flickr is well-defined, but not broadly used. Thus, a second and possibly more important factor in the higher proportion of annotations from the core vocabulary on Last.fm may be that Last.fm facilitates observation of other user's tagging habits (through publicly viewable tag distributions on resources that are not available on Delicious or Flickr). This may allow for the emergence of a large-scale, socially shared vocabulary responsible for most annotations. 

The Delicious data is unique with respect to the slow decay of $\rho$, suggesting that the core vocabulary may be considerably larger. This is consistent with the distribution observed in Figure~\ref{fig:tagFreqs}. Also note the second, earlier peak in $\rho$ for Flickr. This suggests the existence of a smaller subset of popular tags (the top 10) within the larger core vocabulary. In fact, eight of the top 10 tags (``2004'', ``2005'', ``family'', ``friends'', ``japan'', ``party'', ``travel'', and ``wedding'') are the same for $S$ and $\lnot S$. 

Taken together, these results suggest that supertaggers share a core, popular vocabulary with other users, but deviate with respect to the many idiosyncratic and ``singleton'' tags in the data.

\subsection{Differences in Tagged Content}
\label{sec:content}
Having explored aggregate differences in vocabulary, we now ask if the resources tagged by supertaggers differ from those tagged by other users. Supertaggers clearly tend to tag many more items than other users (by at least an order of magnitude, see Tables \ref{tab:folkstats} and \ref{tab:userStats}). Last.fm is particularly notable here, as the total number of items tagged by supertaggers actually exceeds that tagged by other users. Overlap is substantial, however, with 66\% of annotations for Delicious and 78\% for Last.fm occurring for items tagged by both groups. Note that users in Flickr exclusively tag their own photos, so the sets of Flickr items tagged by $S$ and $\lnot S$ are totally disjoint.

Similar to the analyses of tags performed in the previous section, we compare how users' annotations are distributed over item popularity (as measured by total annotation count). However, Figure~\ref{fig:itemFreqs} shows the \emph{cumulative} proportion of tagging in $S$ and $\lnot S$ allocated to items tagged at least a given number of times. 

In contrast to tag usage, we find much more pronounced differences here, both between $S$ and $\lnot S$, and between datasets. On Delicious and Flickr, supertaggers assign proportionally fewer annotations to infrequently-tagged, less popular items (i.e., those with < 10 total annotations). This is is consistent with users generally tagging their own, idiosyncratic content. In contrast, Last.fm supertaggers are proportionally \emph{more} likely to tag less popular content. This suggests that most (non-supertagger) users are more likely to tag shared, popular content on a broad folksonomy like Last.fm.

Note that Flickr shows much greater dominance (across $S$ and $\lnot S$) of singleton and near-singleton tagging. This is confirmed by calculating the Gini coefficient over item annotation counts, which is quite low for Flickr ($0.376$ for both $S$ and $\lnot S$), suggesting high ``equality'' over items (i.e. most items are tagged a similar number of times). The corresponding Gini coefficient for Delicious is much higher (0.703 for $S$, 0.713 for $\lnot S$). This finding is not surprising, though, given that on Flickr users are only tagging their own items. Thus the distribution of tagging over item popularity for Flickr is, in effect, showing the number of times users in $S$ and $\lnot S$ tend to tag each of their uploaded photos.
	
\begin{figure}[t!]
	\centering
	\includegraphics[width=9cm]{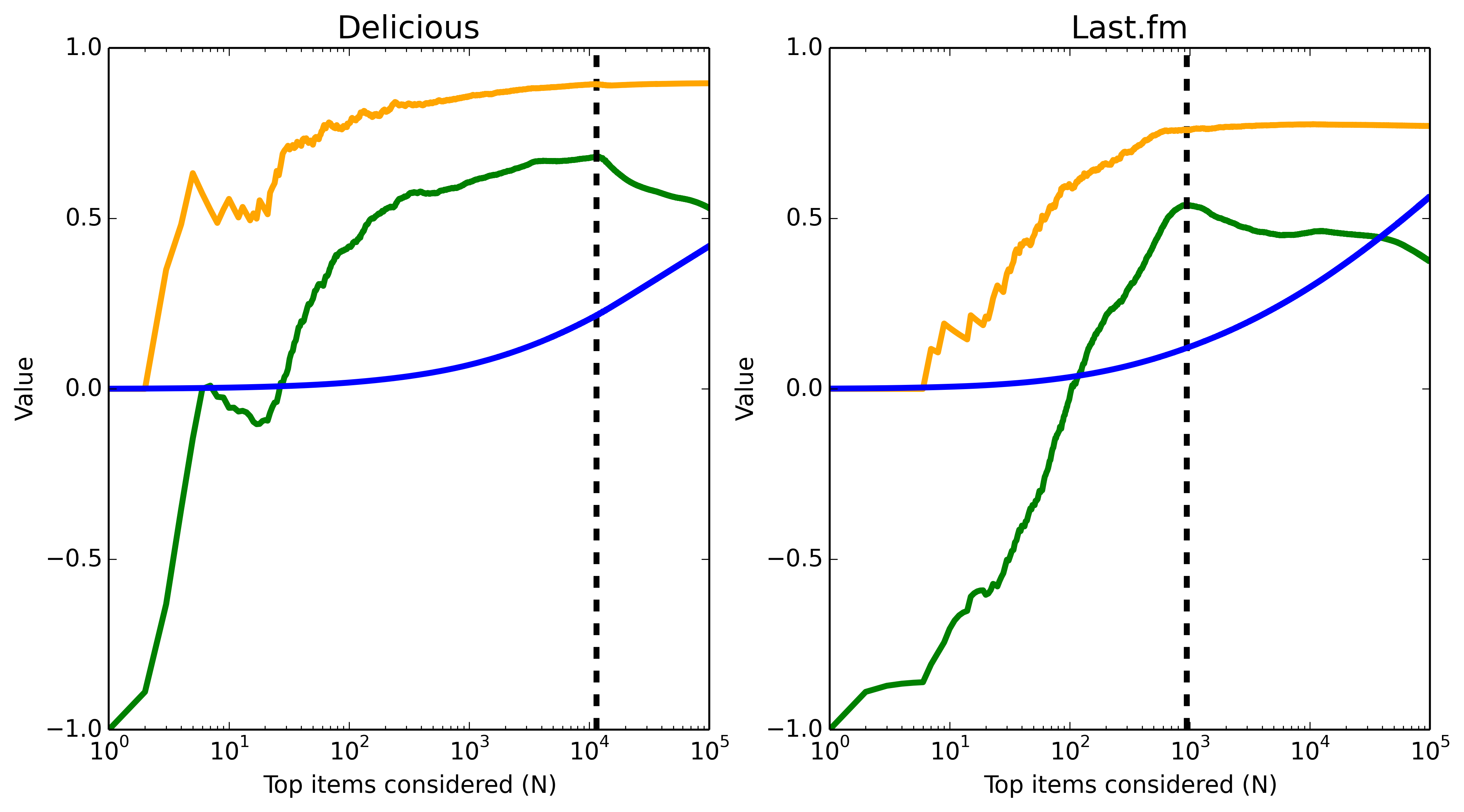}
	\caption{Spearman's $\rho$ and cosine similarity between $S$ and $\lnot S$ as a function of $N$, considering only the top $N$ most popular items overall from each sub-folksonomy. Also plotted is the proportion of all annotations across the full folksonomy assigned to the top $N$ items from both groups. The vertical dashed lines show the maximum values of $\rho$ and occur at $N=11,434$ (Delicious) and $N=943$ (Last.fm).}
	\label{fig:corrsItems}	
	\end{figure}

We also replicate the correlation and similarity analysis from the previous section, but this time comparing the distributions of tagging over items, as opposed to over tags (Figure~\ref{fig:corrsItems}) for the top 1,2,\dots,$N$ items, following the procedure illustrated in Figure~\ref{fig:corrsTags}. From these results, we can conclude the following:  First, the ``core'' set of tagged items, as operationalized by the peak in Spearman's $\rho$ as in the previous analyses, is much larger for items than it is for tags (over 11,000 items for Delicious, and close to 1,000 for Last.fm). Second, although the cosine similarity of item vectors stabilizes at relatively high values (0.89 for Delicious, 0.76 for Last.fm), these similarities, as well as the corresponding peaks in rank correlation (0.681 versus for Delicious, and 0.540 for Last.fm), are substantially lower than the corresponding values for the tag distributions. Finally, the proportion of total tagging activity from the core set of items is lower than that from the core tags (21\% for Delicious and 12\% for Last.fm).

\begin{figure}[t!]
	\centering
	\includegraphics[width=9cm]{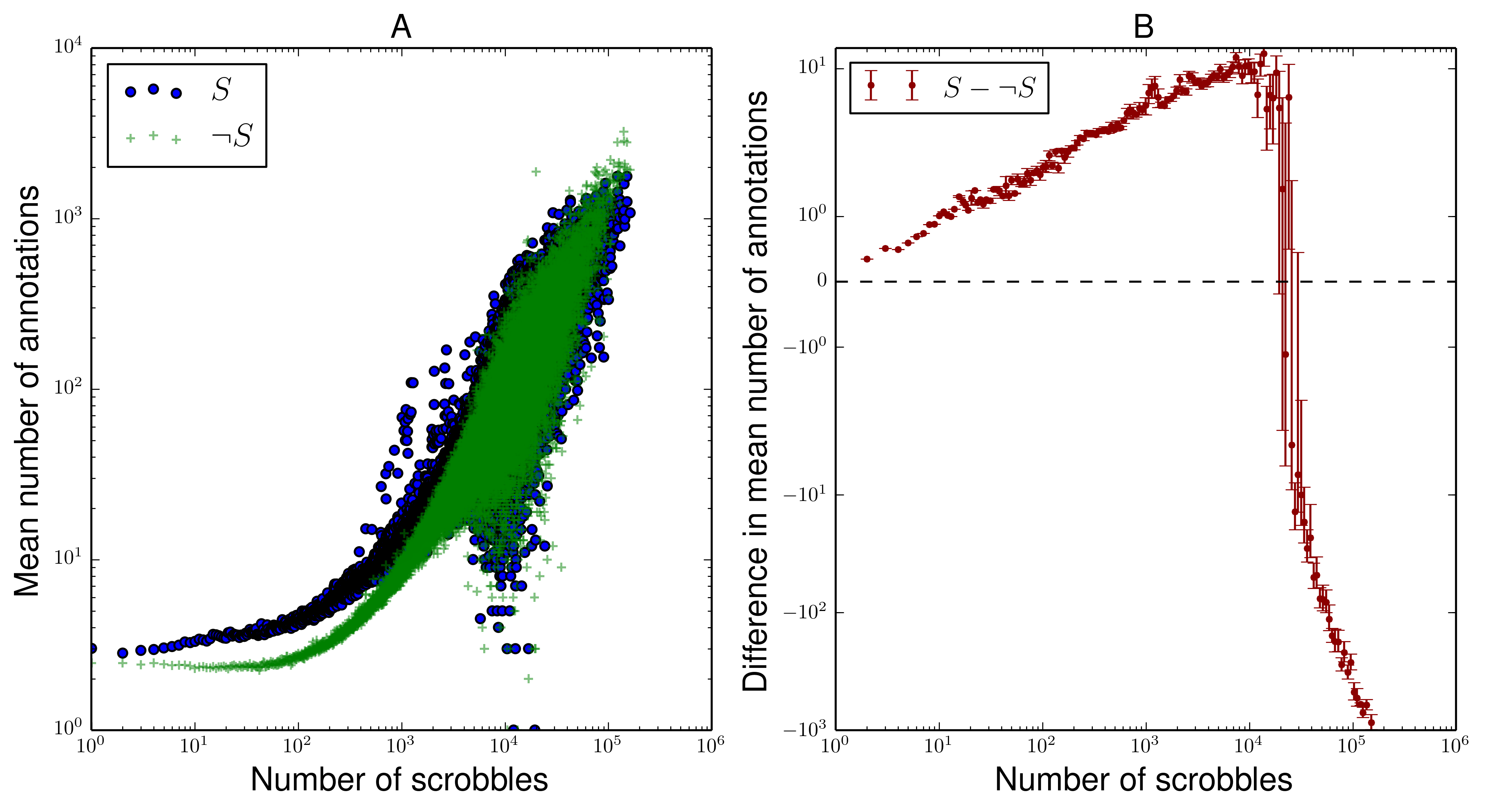}
	\caption{Mean number of annotations by $S$ and $\lnot S$ on Last.fm for items with a given \emph{global} scrobble count (A), and difference in mean number of annotations between $S$ and $\lnot S$ (B). Differences in B are plotted as function of logarithmically-binned scrobble count, and error bars show $\pm 1$ standard error.}
	\label{fig:avs}	
	\end{figure}

\begin{figure}[t!]
	\centering
	\includegraphics[width=9cm]{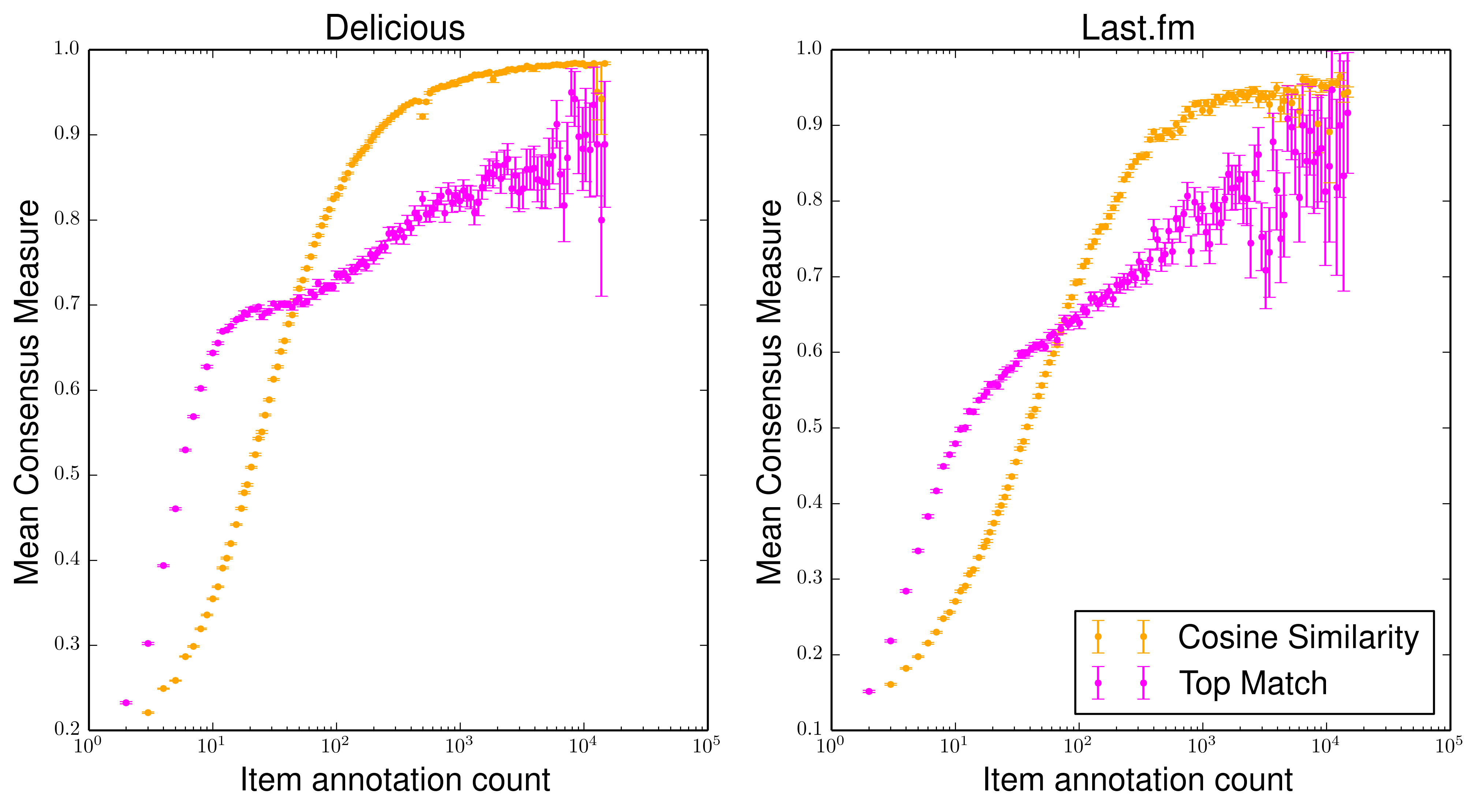}
	\caption{Mean proportion of items on which both groups agree as to the most popular tag (``Top Match'', magenta) and cosine similarity (orange) of tag distributions, as a function of logarithmically-binned annotation count. Error bars show $\pm 1$ standard error.}
	\label{fig:consensus}	
	\end{figure}

Taken together, these results indicate that there is quantifiably less similarity between $S$ and $\lnot S$ with respect to \emph{what} is tagged than which tags are used (for Delicious and Last.fm; again, there is no overlap for Flickr). Therefore, the ``core'' of heavily tagged items is much less clearly defined than the ``core'' of common tags.


One analysis permitted by our supplemental Last.fm data is to compare the popularity of the items tagged by $S$ and $\lnot S$ using an exogenous (i.e. independent of tagging) measure of popularity, namely the global number of scrobbles (listens). In Figure~\ref{fig:avs}A we plot the mean number of annotations in $S$ and $\lnot S$ for items with a particular \emph{global} (i.e. across all users) number of scrobbles.\footnote{This measure is limited to tagged \emph{songs}, not albums or artists.} Though the overall shapes of the distributions are similar, there is a small but reliable effect of supertaggers being more likely to tag items with lower scrobble counts than other users. This is clarified Figure~\ref{fig:avs}B, which shows the difference in mean number of annotations of items as a function of global scrobble count for $S$ and $\lnot S$. Thus, according to an exogenous popularity measure, we find that supertaggers are disproportionately likely to tag less popular content, while non-supertaggers are more likely to tag popular content. This is consistent with the findings with respect to non-exogenous popularity in Figure~\ref{fig:itemFreqs}.

\begin{figure*}[bh!]
	\centering
	\includegraphics[width=18cm]{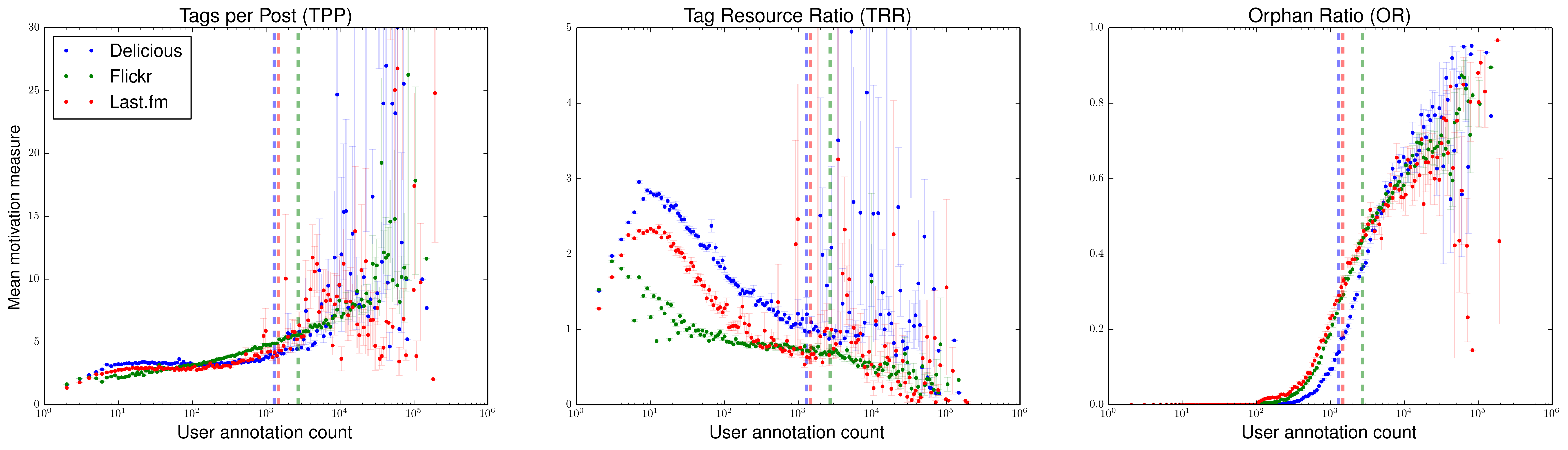}
	\caption{Mean categorizer/describer measures from \citet{Korner2010a} as a function of logarithmically binned annotation count. Shown are Tags Per Post (TPP), Tag-Resource Ratio (TRR), and Orphan Ratio (OR). Error bars show $\pm 1$ standard error. Vertical dashed lines show the supertagger/non-supertagger thresholds.}
	\label{fig:catDesc}	
\end{figure*}
\subsection{Consensus Effects}
\label{sec:consensus}
Having explored aggregate-level differences in items tagged and vocabulary used, it is reasonable to ask -- for those items tagged in both $S$ and $\lnot S$ -- whether or not supertaggers agree with other users as to how particular items ought to be tagged.  Various existing work has established that tagged resources tend to show consensus effects as they accumulate annotations, as measured by a stabilization of the relative proportions of different tags assigned \citep{Golder2006,Robu2009}. Our approach is different however, as we are curious if there is consensus between the two different groups of taggers we have defined for those items tagged by both groups. 

We present two formulations of consensus here. The first, and simplest, measures whether or not the most popular tag for a given item in $S$ is the same as in $\lnot S$. The second measure is the cosine similarity between the distribution of tags assigned to an item in $S$ and in $\lnot S$. These allow us to measure consensus at two levels of granularity, with the first addressing the question of whether users in $S$ and $\lnot S$ agree as to the single ``best'' tag for an item, and the second measuring the overall level of agreement between the two groups. Because we know resources' overall tag distributions tend to stabilize as they accumulate more annotations, we calculate these measures for all items, averaging over items with similar numbers of annotations.\footnote{Because data is sparse for high annotation counts, simply averaging over items with the same annotation count results in a noisy plot that does not make clear the general trends in the data. Thus we bin the data logarithmically (such that the bins for larger annotation counts are wider), and then average over the values within each bin. For this particular plot, the bins are cut at the values $2^{i}, i\in {0.0,0.1,0.2,\dots 14.0}$, and other logarithmically binned plots use a similar procedure (only varying the maximum value of $i$ so as to accurately capture the range of the data).} The results are shown in Figure~\ref{fig:consensus} for Last.fm and Delicious (the analysis is impossible on the disjoint Flickr distributions).
	
The results are consistent with existing work, as items that have been tagged more show, on average, both greater similarity in the distributions of tags assigned to them and in the probability that $S$ and $\lnot S$ will agree as to the top tag. Thus it appears that, as items accumulate more total annotations, $S$ and $\lnot S$ tend to converge as to how those items ought to be tagged. It is notable, however, that on Last.fm the consensus values are uniformly noisier and lower on average than on Delicious. This is true despite the fact that the Last.fm data has a greater percentage of total items tagged by both $S$ and $\lnot S$ (78\% versus 66\% for Delicious). Thus, despite being a broad folksonomy with more users socially tagging shared content, Last.fm demonstrates less pronounced consensus effects.

\section{What makes a supertagger?}
\label{sec:implications}
We have presented the differences in tagging habits between supertaggers and other users, but what might be driving them? Two reasonable questions to ask are (a) do supertaggers' motivations for tagging differ from those of other users, and (b) Are supertaggers ``better'', or more expert, taggers? Here we briefly address these questions quantitatively.

\subsection{Motivational Effects}
\label{sec:motivation}
Our characterization of user motivations follows that of \citet{Korner2010,Korner2010a}, who locate users along the categorizer-describer spectrum. Categorizers are users who constrain their tagging vocabularies to construct personal taxonomies for later browsing; in contrast, describers annotate content freely with a wide assortment of tags to facilitate later keyword-based search. We quantified user motivation along this spectrum using three metrics developed by K\"{o}rner and colleagues: tags per post (TPP), tag/resource ratio (TRR), and the orphan ratio (OR). TPP measures the number of distinct tags a user annotates an item with on average. Based on \citet{Korner2010,Korner2010a}, we expect describers to annotate items with more tags on average, and thus score higher on this measure. TRR is the ratio of the vocabulary size of a user to the total number of items tagged by that user. We expect categorizers to maintain their limited, personal taxonomies in tagging, and thus use fewer unique tags overall, thereby scoring lower on this measure. OR relates the vocabulary size of a user to the number of seldom-used tags for this user (i.e. what proportion of a user's tags are ``orphans''?). We expect describers to be less motivated to reuse tags, and thus score higher on this measure. Though there exist other measures of motivation, we limit our analyses to these three in light of previous research reporting high correlations between TPP, TRR, OR, and other measures, following the recommendation of \citet{Zubiaga2011}. For full details on the calculations of each measure, see \citet{Korner2010,Korner2010a}.

Figure~\ref{fig:catDesc} presents the TPP, TRR, and OR scores as a function of users' total annotation counts. Across all datasets, although the data is unsurprisingly noisy for high annotation counts, TPP scores tend to increase as total annotations increase. This suggests that users in $S$ are not simply annotating more items; rather, they are, on average, annotating any given item with more tags than those in  $\lnot S$. We find a similar trend for OR scores: the number of orphaned tags in the vocabulary of a user increases as a function of that user's total annotation count. These two results suggest that supertaggers are more like describers than are non-supertaggers. 

The trend of decreasing TRR scores as a function of total annotations across all datasets presents as a challenge to this interpretation. As explained above, greater TRR scores are characteristic of describers. We believe the discrepancy can be explained, however, by the growth rate of user vocabularies. \citet{Cattuto2007} report sub-linear growth of user tag vocabularies as compared to the total number of annotated items, perhaps reflecting a saturation point in the number of unique tags a given user will employ. 

In sum, then, the results suggest that supertaggers are better characterized as describers, who are more likely to tag the same items multiple times on average, while drawing from a more diverse set of tags. Non-supertaggers, on the other hand, are better characterized as categorizers, maintaining smaller, more structured tagsets. This suggests that supertaggers and non-supertaggers differ with respect to motivations in tagging. This may be reflective of differences in approaches to item retrieval, with supertaggers annotating with more tags for easier lookup. There are of course other approaches to studying tagging motivation, and further work is needed to establish what factors might drive certain users to tag far more than others.

\subsection{Are Supertaggers Expert Taggers?}
A second feature by which to characterize the users of a folksonomy is their expertise, or quality of information provided in their tagging contributions. There is not, however, a single agreed-upon definition of expertise with respect to tagging. As such, we quantify expertise using three approaches. The first, SPEAR, is an established measure that determines expertise through two principles, mutual reinforcement and time of tagging (i.e. expert taggers tend to tag things earlier). We introduce a new measure that, in contrast to SPEAR's tag-level approach, evaluates users on item-by-item basis, assigning higher scores to users annotating an item in agreement with the consensus of annotations for that item. A third measure, inspired by longstanding psychological research, defines expertise as the increased usage of sub-ordinate terms in semantic taxonomies. These approaches are detailed below.
\label{sec:expertise}
\begin{figure}[bh!]
	\centering
	\includegraphics[width=8cm]{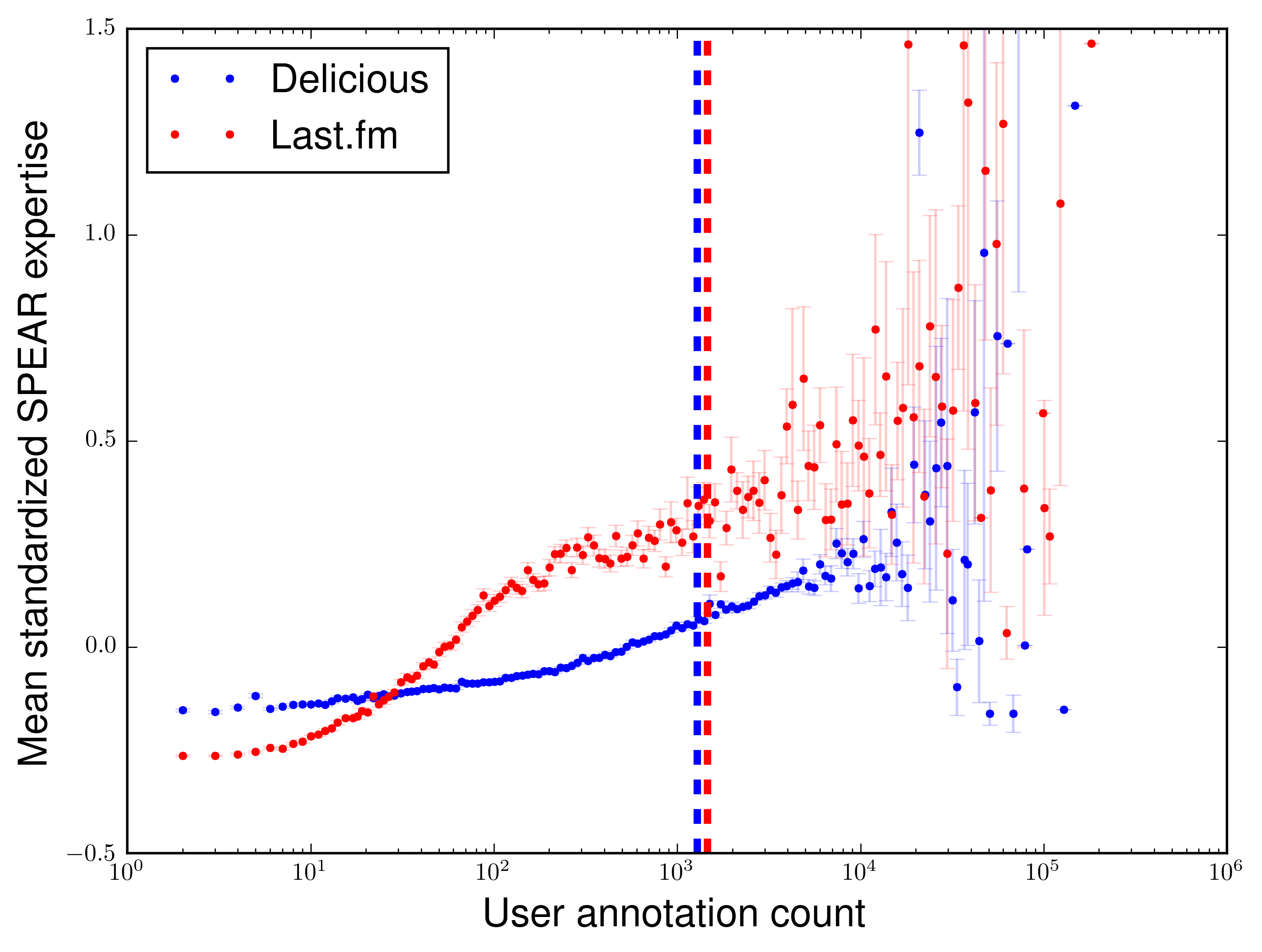}
	\caption{Users' mean standardized SPEAR expertise scores as a function of logarithmically binned annotation count. Vertical dashed lines show the supertagger thresholds for each dataset. Error bars show $\pm 1$ standard error.}
	\label{fig:spear}
\end{figure}
\subsubsection{SPEAR Expertise}
As a first approach to measuring expertise, we used an established measure, the Spamming-resistant Expertise Analysis and Ranking (SPEAR) algorithm. SPEAR has two core mechanisms. First, it is a mutual reinforcement model based on the HITS ranking algorithm \citep{Kleinberg1999}, in which user expertise in a topic (as defined by a particular tag) is based on the quality of the items tagged, and an item's quality is in turn based on the expertise of the users tagging it. Second, it incorporates a discoverer/follower mechanic by assuming that the first users to annotate an item with a particular tag are better at identifying high quality items and are more likely to be experts than those annotating after them. Thus SPEAR is able to rank users in terms of their expertise (or authority) in a topic (tag), favoring those users who are among the first to ``discover'' an item by tagging it. In other words, expertise in a tag is quantified such that users who are among the first to annotate high quality items with that tag are assigned the highest scores. For full details of the algorithm, see \citet{Yeung2009,Yeung2011}. We used the default parameters of the algorithm in the results presented here.

It is important to emphasize that SPEAR is by design a measure of domain expertise in that it provides user expertise scores for a particular tag. That is, each user receives one score per tag that is independent of her scores for all other tags. SPEAR's mutual reinforcement model does not result in the score for each of a user's tags being on the same scale, making the computation of an overall expertise score something of a challenge. As an attempt to address this, we standardized all scores corresponding to a given tag to mean = 0 and standard deviation = 1 (z-score), relative to the distributions of scores for that tag across all individuals using it. Though this allows for a mean score per user, we reemphasize that this is not part of the original intentions for SPEAR. 

For our SPEAR analyses, we used a subset of each dataset corresponding to the top 10,000 most popular tags overall that have at least 10 unique users. This decision was made in part due to computational limitations on calculating scores for all tags,\footnote{Exploratory analyses with various thresholds, however, yielded qualitatively similar results.} and also because SPEAR generates unreliable scores for tags that are very rarely used.\footnote{SPEAR strongly rewards users for being among the first to use a given tag, so for tags used by only one or a small number of users, the algorithm necessarily leads to radically inflated, uninformative expertise scores.} Despite including only a small proportion of total unique tags, this trimming still gives us a reasonable coverage of the annotation data; the top 10,000 tags account for approximately 82.6\% of annotations from Last.fm and 83.6\% from Delicious. Note that the Flickr dataset was excluded from this analysis as its design feature of exclusive self-tagging prohibits the mutual reinforcement necessary for SPEAR. Further, these calculations are performed over the full folksonomy for each dataset (i.e. not considering $S$ and $\lnot S$ independently). Figure~\ref{fig:spear} presents average user SPEAR expertise scores as a function of user annotation count. Consistent with our previous work \citep{Lorince2014}, we find an overall positive relationship in both Last.fm and Delicious such that average user expertise increases with number of annotations, although the data is noisy for high annotation counts. This suggests that supertaggers are not only more prolific, but also more expert, in their tagging behavior, at least as defined by SPEAR.

Despite similar overall trends, there are noteworthy discrepancies in the shapes of the SPEAR score distributions between Delicious and Last.fm. Last.fm exhibits an earlier peak in the growth of user expertise scores as total annotations increase, and scores remain consistently higher afterwards. Interpreting this finding, however, is complicated by the fact that SPEAR was not intended to be used as a measure of overall user expertise in the way we have presented. Though we took additional steps to make the user expertise scores across tags comparable (via the score standardization), we have still used SPEAR in a non-traditional manner and hesitate to make strong claims as to what may be driving the observed differences between datasets. A further complication  arises as a result of the low temporal resolution of our Last.fm data, which makes it difficult to determine the true sequence in which tags were assigned to an item (because we only know the month in which an annotation was made, all annotations in the same month are necessarily treated as having been generated simultaneously). This issue is not relevant to the Delicious dataset.

\subsubsection{Consensus-Based Expertise}
Given these issues, we developed two novel measures of expertise targeted specifically at providing general expertise scores for each user. Our measures capture two contrasting, but reasonable characterizations of expertise. The first, described in this section, assumes that an expert user will, on average, be in agreement with the consensus on how an item should be tagged. In other words, an expert should be in alignment with the ``wisdom of the crowd'', being more likely to assign the ``correct'' tag to a given item. Though SPEAR indirectly captures consensus through mutual reinforcement, our new measure explicitly does so by computing the popularity of a user's tag choice in annotating an item relative to the most popular tag for that item over the entire lifetime of the item, for each annotation the user generates. Thus, whereas SPEAR calculates expertise over tags, our measure generates a raw score for each of a user's annotations relative to the item tagged, without reference to how that tag is used globally. To calculate a user's overall average expertise score, the score for each of a user's annotations is weighted by the logarithm of the total annotation count for the item tagged. In this way, tagging items with very low annotation counts contributes little to a user's overall score, while annotations of heavily tagged items contribute more. This captures the intuition that the more total times an item has been tagged, the better defined the tagging consensus for that item is. Note that our novel measure also does not take time into consideration. For our measure, it matters less when a user annotates an item than it does that her tagging choice coincides with the eventual consensus of other users for that item.

The measure is formally defined as follows. For each annotation (user-item-tag triple, ignoring time) we calculate a raw expertise score, $E_{u,i,t}$:
\begin{equation*}
	E_{u,i,t} = \frac{F(t,i)-1}{\max(F(x,i),x\in T(i))}
\end{equation*}
\noindent where the expertise score assigned to a user $u$ annotating item $i$ with tag $t$ is the frequency of that tag, $F(t,i)$, in the overall distribution for item $i$, divided by the frequency of the most popular tag for that item ($\max(F(x,i),x\in T(i))$, where $T(i)$ is the set of all tags assigned to item $i$). As this is a consensus-based metric, we ignore a user's own contributions to that tag distribution by subtracting 1 from the numerator.\footnote{Though it is not shown here, in the event that the user has tagged an item with the most popular tag, we assign a consensus score of 1.} To determine a user's average expertise score over all his or her annotations, $\bar{E}_{u}$, we calculate the following weighted mean:
\begin{equation*}
	\bar{E}_{u} = \frac{\sum_{i,t}{E_{u,i,t}W_{u,i,t}}}{\sum_{i,t}{W_{u,i,t}}}, W_{u,i,t}=\log_{10} \left[\left( \sum_{t=1}^{n}F(t,i) \right) - F(u,i) \right]
\end{equation*}
\noindent That is, we calculate a weighted mean of the expertise score for each of a user's annotations, where the weight, $W$, is equal to the the logarithm of the \emph{total} number of annotations across all users for the item tagged ($\sum_{t=1}^{n} F(t,i)$), minus the current user's contribution to that distribution (i.e. the total number of times that user tagged the item, $F(u,i)$). In cases where a user has assigned multiple tags to the same item, we only include the single highest expertise score for that item in the user's mean score. In this way, we capture whether or not the user knows the ``best'' tag for an item, without penalizing her if she additionally assigns other tags to it. This, combined with the low weighting of items tagged only a few times, sidesteps the issue that supertaggers tend to use many idiosyncratic tags (which necessarily are not ``expert'' tags under any consensus-based measure). In effect, this analysis allows us to determine whether supertaggers show expertise with relatively popular tags, while not considering their usage of idiosyncratic tags.\footnote{Additionally, we ran exploratory analyses in which we only considered tags with at least 10 unique users, thereby calculating expertise scores for only non-idiosyncratic tag use. The results were qualitatively similar.} For the same reason as above, Flickr was excluded from this analysis. 

\begin{figure}[ht]
	\centering
	\includegraphics[width=8cm]{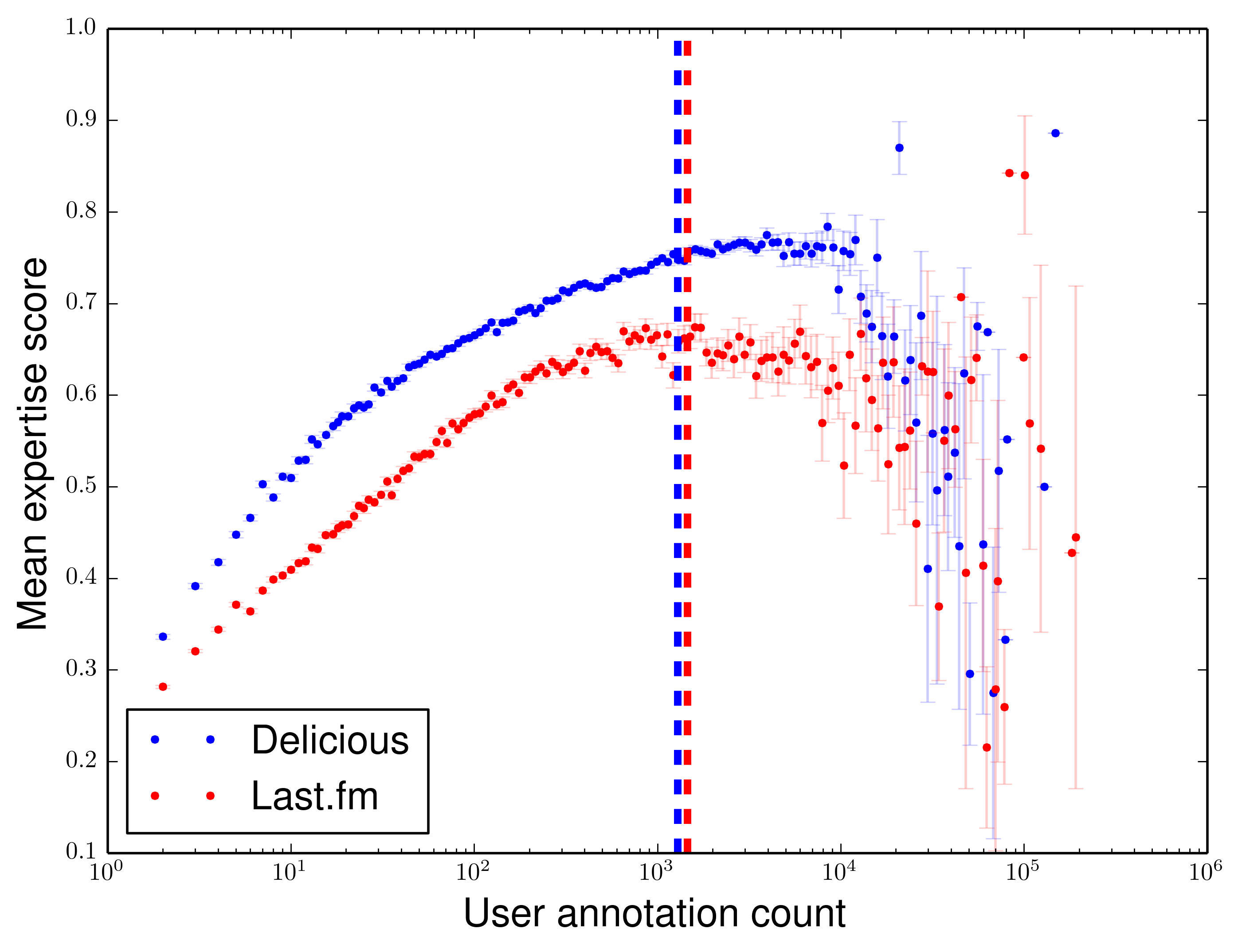}
	\caption{Users' mean consensus-based expertise scores as a function of logarithmically binned annotation count. The vertical dashed lines show the supertagger thresholds for each dataset. Error bars show $\pm 1$ standard error.}
	\label{fig:lorinceTrim}
\end{figure}

Figure~\ref{fig:lorinceTrim} presents mean consensus-based user expertise as a function of user annotation count. In contrast to SPEAR, and somewhat surprisingly, the results show an inverse-u shape. Expertise scores increase monotonically as a function of annotation count for $\lnot S$ (left of the dashed lines in Figure~\ref{fig:lorinceTrim}); the growth of expertise scores, however, tapers off for $S$ before decreasing substantially for the most prolific taggers. Expertise, then, as defined by agreement with the consensus, increases with the number of users annotations for all but the most prolific of taggers.

Of note, however, is the high variability in the consensus-based expertise scores of these most prolific taggers. Though this is in part due to a rapidly diminishing sample size of users as total annotation count increases, it may also be indicative of divergent tagging away from the consensus in a subset of users in $S$. Whether this is a legitimate difference in tagging behavior, an artifact of spammers, or simply noise is beyond the scope of this paper, but again emphasizes the importance of investigating separately the behaviors of supertaggers and non-supertaggers. 

Of additional note is the clear difference between datasets for users with similar numbers of annotations in Figure~\ref{fig:lorinceTrim}. Users in Delicious show reliably higher expertise scores as compared to similarly prolific users on Last.fm. This result is not entirely surprising, however, given the results discussed in Section~\ref{sec:consensus}. We found that supertaggers and non-supertaggers showed greater agreement as to how items ought to be tagged on Delicious than Last.fm, so it is little surprise that, on average, consensus based expertise scores are higher for Delicious. Exactly what is driving this is less clear, however. Both Delicious and Last.fm provide frequency-based tag recommendations (i.e. suggesting the top five most popular tags for an item in the tagging interface) that presumably would encourage consensus effects, but Last.fm is unique in that users can choose to explore the full tag distributions for an item and thereby be exposed to more tags. Other factors beyond the scope of this paper are certainly at play, both social (e.g. do the publicly shared tag distributions for items somehow encourage greater tagging diversity?) and content-based (is music inherently more difficult to reliably classify than are webpages?), and deserve attention in future work.

\subsubsection{Term-Depth Expertise}
An alternative, more psychologically grounded approach to measuring expertise is one based on classic research \citep{Rogers2007,Rosch1976} showing that people tend to prefer basic level categories to describe objects, whereas domain experts are more likely to use subordinate labels to describe objects within in their area of expertise. For example, non-experts may refer simply to a ``tree'' (a basic level category), while a botanist is likely to identify that same tree at a lower level (e.g. ``spruce'', a sub-ordinate category). Neither group is likely to refer to it simply as a ``plant'' (a super-ordinate category). Applying this to our analyses, if supertaggers demonstrate more expertise than other users, we should expect from them to use more sub-ordinate terms on average than other users.


To test for this, we employ methods developed by \cite{Kubek2010} to create a tagging taxonomy for each folksonomy. Their algorithm measures the conditional probabilities of tag co-occurrence over items to infer when one tag is a sub-class another. For example, the term ``classic rock'' is more likely to co-occur with the term ``rock'' than vice-versa, and is therefore likely to be a sub-class of ``rock''. Full details of the algorithm appear in the original study, but it involves first calculating all pairwise conditional probabilities between tags (i.e. both $P(A|B)$ and $P(B|A)$), then, following thresholds defined in \cite{Kubek2010}, defining the sets of sub- and super-classes for each tag. From these values, a taxonomy defining the hierarchal relationships between all tags can be extracted. Once this taxonomy is defined, we can use a given tag's depth in the tree as a proxy for how sub- or super-ordinate of a term it is (i.e. tags at root nodes are presumably super-ordinate terms, while tags further down the tree are sub-ordinate terms).

Because the method involves calculating all pairwise similarities between tags and is thus computationally intensive, and also because rarely used tags are unlikely to co-occur across enough items to effectively determine their associated conditional probabilities, we limit the analysis to the subset of tags used in the above SPEAR analyses (the top 10,000 tags with at least 10 unique users across the full folksonomy). For this measure, however, Flickr data can be analyzed. As mentioned before, however, this gives us good coverage of annotations across datasets (82.6\% for Last.fm, 83.6\% for Delicious, and 67.6\% for Flickr). The analysis results in a forest of taxonomies per folksonomy, as not all tags fall under a single root node. Each tag was part of one of these taxonomies or else a disconnected node. We filtered  our results to exclude disconnected nodes, as these are tags without discernible relationships to other tags, and thus without well-defined taxonomy depth. This reduced our set of considered tags from 10,000 to 4,724 for Last.fm (covering 68.1\% of all annotations), 2,224 for Flickr (29.3\%), and 5,148 for Delicious (44.1\%). Each tag is assigned a simple depth score based on the taxonomy (i.e. a root node has a score of zero, its children have a score of one, its children's children have a score of two, and so on). Furthermore, we normalize these raw scores, which ranged from zero to a maximum depth of five, by dividing them by the maximum depth of the branch the node was contained in, thus normalizing to a 0-1 scale. This normalized score is our measure of term-depth expertise, with the assumption that leaf nodes are more specific than root nodes.
\begin{figure}[ht] 
    \centering
    \includegraphics[width=8cm]{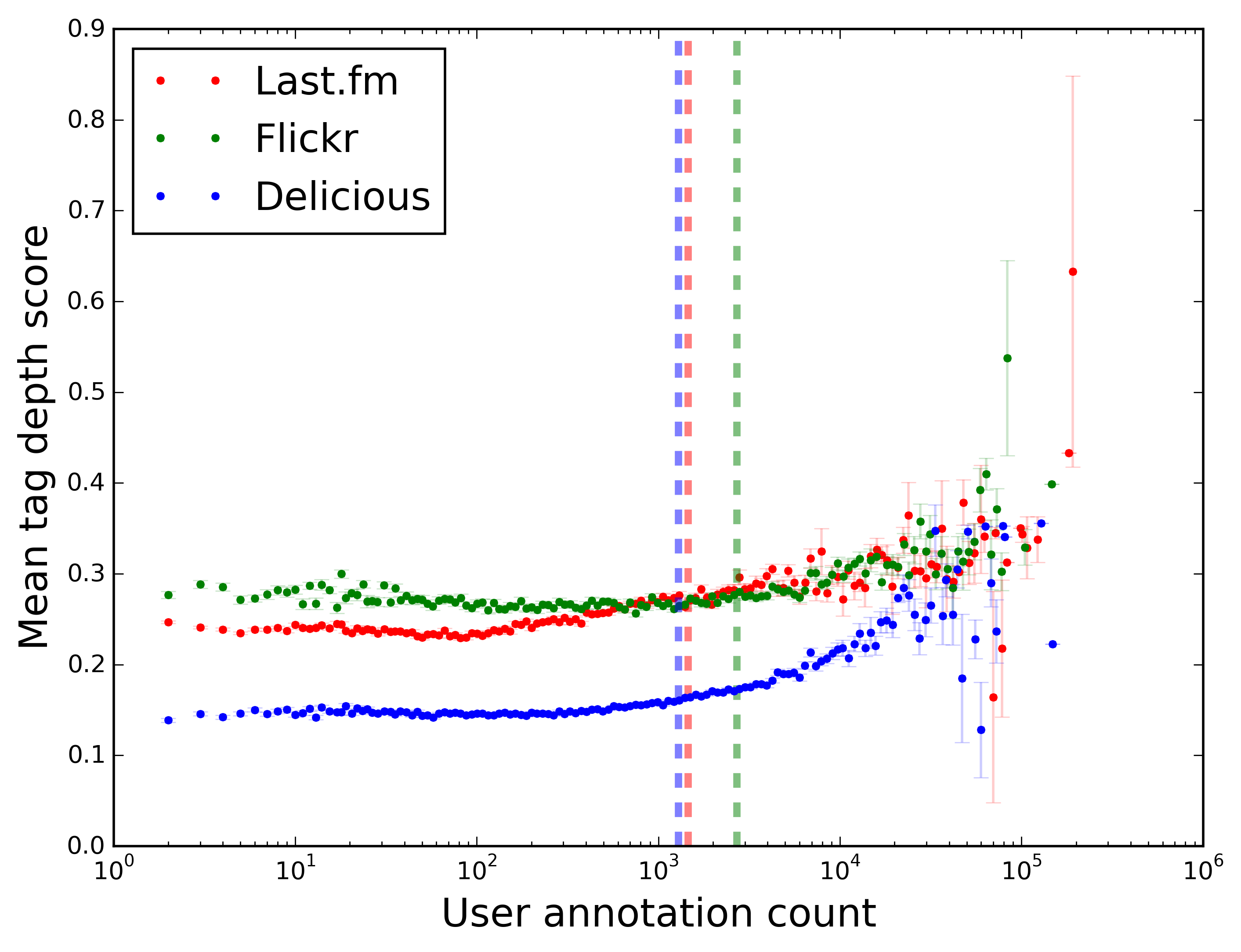}
    \caption{Users' \emph{vocabulary-level} mean term-depth expertise scores as a function of logarithmically binned annotation count. The vertical dashed lines show the supertagger thresholds for each dataset. Error bars show $\pm 1$ standard error.}
    \label{fig:newExpertise_vocab}
\end{figure}

After mapping each tag to its depth score, a user's average depth-score expertise is simply the average over annotations (i.e. over each instance of using a tag) of tag depth scores. In contrast to our other expertise measures, we do not observe systematic increases in expertise as annotation counts increase, and generally speaking there is no substantial difference in expertise between supertaggers and non-supertaggers (hence we have not included these results visually). We repeated the analysis, however, at the vocabulary level, which means we averaged each user's tag depth scores over the unique tags she employs, regardless of how many times each tag was used. These results, presented in Figure~\ref{fig:newExpertise_vocab}, show a small but clear effect of increasing depth scores for the most prolific users. 

Thus we can conclude that, at least for the tags we were able to examine with this method, the most prolific taggers tend to have vocabularies consisting of more subordinate terms as compared to other users. This trend does not extend to their aggregated tagging activity -- that is, when averaging depth scores across all of a users' annotations, rather than across only unique tags -- as we observed with the previous two expertise measures. The method is also limited in that we can only make claims about a proportion of unique tags in the dataset. Future work may be able to use an information theoretic approach to compare the specificity of tags for the purposes of expertise analysis, and thereby improve on the method used here. Nonetheless, these findings are suggestive that supertaggers introduce more terms associated with greater expertise into the folksonomy than other users, even if they do not make up a large proportion of their tagging activity. 

\section{Discussion and Conclusions}
\label{sec:conclusion}

The principal contributions of this work are the following:
\begin{itemize}
	\item A formalization of the disproportionate contribution by ``supertaggers'' to a folksonomy;
	\item an analysis of the differences between these taggers and their non-prolific counterparts, at the levels of the users themselves and the folksonomic structures they generate; and
	\item an analysis of the role of expertise and tagging motivation in these differences, including novel metrics for tagging expertise.
\end{itemize}

Our results demonstrate that the most prolific taggers are not simply generating a greater volume of annotations in a manner consistent with ``the crowd''. Instead, their tagging patterns are quantifiably different from those of other users across datasets. With respect to tag vocabulary, we find that both groups use many of the same most popular tags, but disagree on the long tail of less common tags. With respect to items tagged, results differed across the datasets. On Last.fm, supertaggers allocate proportionally more annotations to less popular items than do other users, while on Flickr and Delicious the opposite trend held. On Last.fm, this suggests that the tagging of users in $S$ is more exploratory, disproportionately tagging content in the long tail of obscure, unpopular items. This was confirmed by an exogenous measure of item popularity, as well (see Figure~\ref{fig:avs}).

We hypothesize that these differences are driven by the different interaction paradigms of broad and narrow folksonomies. In narrow folksonomies, such as Delicious and Flickr, users are predominantly uploading and tagging their own content (photos and bookmarks). Distributions of tagging over item popularity are thus driven at least partly  by the number of times users tend to tag any given item. In broad folksonomies, such as Last.fm, shared items are publicly tagged by multiple users, precisely the process that ostensibly allows for the ``wisdom of the crowds'' to emerge in collective classification. However, we found that users demonstrate \emph{less} consensus about what is tagged and what tags are used than on Delicious or Flickr. On Delicious, multiple users may save and tag the same bookmarks, but presumably tag privately for their own organization of bookmarks. Thus one could reasonably expect lower consensus across users as to how those items should be tagged in contrast to Last.fm, where users knowingly tag publicly-shared content, but we found exactly the opposite. This surprising result demands further work exploring precisely what factors make consensus more or less likely in collaborative tagging.

Supertaggers also differed in their scores on previously established measures of user motivation. Supertaggers are not only annotating more often but also using broader tag vocabularies and assigning more unique tags to each item. These results suggest not only that supertaggers are better characterized as describers than categorizers, but also that they may differ in the underlying reasons for their patterns in tagging, perhaps reflective of differences information retrieval strategies. 

We also found that average user expertise increases as a function of number of annotations made by the user, at least for the majority of users. This result was consistently found across two measures of expertise, the SPEAR algorithm and our novel measure, both of which define expertise in part based on consensus in tagging with the majority of users. A third measure of expertise, based on the specificity of terms used (i.e. sub-ordinate versus super-ordinate), did not show a similarly clear relationship between amount of tagging and expertise, but does suggest supertaggers are more likely to have more expert tags in their vocabularies than other users. Across all three measures, we found expertise scores vary considerably for the most active of users. This was especially true for our consensus-based expertise measure, where expertise scores exhibited an overall decline for the most prolific taggers. Due to small numbers of the most active users it is difficult to discern if this finding is reflective of a discordant subset of supertaggers or simply a result of noise. Thus, refinements of expertise measurement in tagging remain a useful avenue future work. 

Of course, expertise can be defined in many ways. Two of the measures used here assume a connection of expertise to consensus. Implicit in this, however, is a further assumption that the crowd will invariably arrive at an accurate description of an item. This may not be so, especially in the case of broad folksonomies where social interaction may have an important effect on the formation of tag distributions across items. For example, the first annotations of an item may be the most important to the resulting consensus of tags if social imitation is at work. If supertaggers are more often than not among the first users to annotate an item, as our SPEAR results suggest,\footnote{This is confirmed directly in the case of Delicious by supplemental analyses (not presented here) examining only the average point at which supertaggers annotate items relative to other users. As expected, more prolific taggers do tend to tag items earlier than others. In the case of Last.fm, however, the low temporal resolution of our data makes the results of such an analysis unreliable and difficult to interpret.} then it is possible that their early contributions may shape the resulting distributions in ways favorable to higher expertise scores as defined by our measures. Especially given that our non-consensus-based expertise measure did not show the same pattern as the other measures, our results suggest the need for exogenous measures of expertise. This can be explored in future work by defining a user's expertise in a musical genre based on her listening habits, and then exploring how this interacts with her patterns of tagging.

While tagging has been hailed as an example of the ``wisdom of the crowd'', we have shown that the majority of tagging is not done by ``the crowd'' at all. These results call for questioning just how much ``collective classification'' is actually happening in social tagging systems. While the most popular items are tagged by many users, the long tail of less popular items are being tagged mostly by supertaggers, especially in broad folksonomies like Last.fm. This is not necessarily an argument against folksonomies, especially if supertaggers are shown to be experts. We only claim that when designing or studying social tagging systems we need to be sensitive to not only variation in how much people tag, but the varying manners in which they do so. Determining whether the ``division of labor'' we see among taggers serves to generate a more (or less) usable semantic structure than would be created by users with more homogenized tagging strategies is a promising direction for future research. 

There are of course limitations to the methods we have used here. Most notable is our arbitrary partitioning of users into supertaggers and non-supertaggers (by splitting the data in half on total annotations), especially given that none of our results provide evidence of a strong qualitative division of users into these two groups. To be clear, we do not make the claim that there is anything inherently meaningful about the particular threshold we used to define supertaggers. Rather, we used this division of users to demonstrate (a) the extreme skew in tagging contributions towards prolific taggers, and (b) that users at one end of the annotation spectrum tag quantifiably differently from users at the other end. But our division of users means that it could be the case, for instance, that the measured differences in motivation of supertaggers are partly a function of their tagging more obscure items. This might occur if more obscure items do not fit canonical musical categories and demand multiple classifications such that users tagging them appear more like describers than categorizers, even when this does not reflect a fundamental motivational difference. Relatedly, the motivations of supertaggers may not reflect internal, stable user traits but may instead result from interacting with the folksonomy over time. By virtue of discovering more obscure items through increasing use, users' motivations may transition from resembling categorizer to describer behavior for the reasons described above. In fact, we do not yet understand whether any of the observed differences between supertaggers and non-supertaggers reflect anything inherently different about users, or might predictably emerge as users tag more. Much future work thus remains to be done to understand how tagging patterns evolve over time. There also is the question of the extent to which our results were affected by spam tagging, which we did not directly address here. Effective identification and elimination of prolific spam taggers might shift the dominance in annotation counts away from the most prolific taggers.

Finally, our analyses do not account for within-user variation, which may turn out to be crucial to this kind of work. For example, we generated a single, overall expertise score for every user, even though that may not be appropriate. Users presumably show varying levels of expertise in different domains, such that the average across those domains may not be meaningful. A similar case might be made for the simple classification of users by annotation count. A user who tags many items within a single topical domain is behaving differently from one who tags the same number across a broad variety of domains.

A major question left unanswered in this work is \emph{why} the differences we have observed exist. Why do some users become supertaggers, while others tag very little at all? We can only speculate at this point, but our results are suggestive of some possible explanations. First, differences in tagging motivations, as measured by K\"{o}rner and colleagues' methods, suggest that supertaggers behave more like describers than categorizers, so it is possible that describer-like tendencies encourage users to tag more (though, of course, the reason for these tendencies remain unknown). Second, differences in expertise are suggestive that more expert users tend to tag more, but it remains unknown unknown which way the causality runs: Is greater expertise a result of more tagging, or increased tagging rates a result of greater expertise? Studying trends in expertise over time within users could help shed light on this question.

Social tagging systems represent intriguing environments in which a subset of users are highly active in the absence of clear incentives for doing so (there are no explicit rewards for being a supertagger, social or otherwise). In some cases spamming may be at play, but it is doubtful that this accounts for all or even most cases of supertagging observed in our data. Though we are unable to answer the question of \emph{why} such pronounced differences in tagging activity exist, we believe our analysis of \emph{how} prolific and non-prolific taggers differ represents a substantial contribution to understanding such systems.

Despite the need for further investigation, our work nevertheless presents compelling evidence that the bulk of tagging activity comes from a minority of users whose tagging behavior is quantifiably distinct from that of other users. Thus, it is important for both researchers and designers of collaborative tagging systems to identify and differentially interpret the metadata generated by these supertaggers in order to understand and promote the use of these systems by all.

\end{document}